\documentclass[twocolumn,accepted=2019-07-22,letterpaper]{quantumarticle}
\pdfoutput=1
\usepackage{amsmath}
\usepackage{amssymb}
\usepackage{graphicx}
\usepackage{bm}
\usepackage{physics}
\usepackage[colorlinks=true]{hyperref}
\usepackage[numbers,sort&compress]{natbib}
\usepackage{tikz}
\usepackage{tikz-cd}
\usetikzlibrary{arrows.meta}
\usetikzlibrary{decorations.markings}
\tikzset{middlearrow/.style={
        decoration={markings,
            mark= at position 0.6 with {\arrow{#1}} ,
        },
        postaction={decorate}
    }
}
\DeclareMathOperator\arctanh{arctanh}

\newtheorem{theorem}{Theorem}

\newtheorem{lemma}{Lemma}[theorem]
\definecolor{dualblue}{RGB}{3,101,192}

\begin{document}
\title{Universal logical gates with constant overhead: instantaneous Dehn twists for hyperbolic quantum codes}
\author{Ali Lavasani}
\author{Guanyu Zhu}
\thanks{Current affiliation is \textit{IBM T.J. Watson Research Center, Yorktown Heights, NY 10598}.}
\author{Maissam Barkeshli}
\affiliation{Department of Physics, Condensed Matter Theory Center, University of Maryland, College Park, Maryland 20742, USA
and Joint Quantum Institute, University of Maryland, College Park, Maryland 20742, USA}
\begin{abstract}
A basic question in the theory of fault-tolerant quantum computation is to understand the fundamental resource costs for performing
a universal logical set of gates on encoded qubits to arbitrary accuracy. Here we consider qubits encoded with constant space
overhead (i.e. finite encoding rate) in the limit of arbitrarily large code distance $d$ through the use of topological codes associated
to triangulations of hyperbolic surfaces. We introduce explicit protocols to demonstrate how Dehn twists of the hyperbolic surface
can be implemented on the code through constant depth unitary circuits, without increasing the space overhead. The circuit
for a given Dehn twist consists of a permutation of physical qubits, followed by a constant depth local unitary circuit, where locality
here is defined with respect to a hyperbolic metric that defines the code. Applying our results to the hyperbolic Fibonacci Turaev-Viro
code implies the possibility of applying universal logical gate sets on encoded qubits through constant depth unitary circuits and with
constant space overhead. Our circuits are inherently protected from errors as they map local operators to local operators while changing
the size of their support by at most a constant factor; in the presence of noisy syndrome measurements, our results suggest the possibility of universal fault tolerant
quantum computation with constant space overhead and time overhead of $\mathcal{O}(d/\log d)$. For quantum circuits
that allow parallel gate operations, this yields the optimal scaling of space-time overhead known to date.
\end{abstract}
\maketitle
\section{Introduction}

The significant effects of decoherence on quantum systems require that a fault-tolerant quantum computer
appropriately encode logical quantum information and furthermore apply logical gates directly on the
encoded qubits \cite{nielsen_chuang_2010,Terhal:2015ks}. However it is currently an open question to
understand the ultimate asymptotic resource costs required for performing quantum error correction and
fault-tolerant quantum computation. In this work we improve upon the known optimal asymptotic space-time resource costs by proposing constant depth topologically protected unitary circuits that implement a universal set of logical gates for the Tuarev-Viro codes\cite{Koenig:2010do} defined on hyperbolic surfaces. Our protocols demonstrate that protected universal gate sets can be applied in parallel and with constant space overhead using constant depth unitary circuits.

A quantum error correcting code (QECC) encodes $k$ logical qubits using $n$ physical qubits, such that a logical
error occurs if and only if errors occur on at least $d/2$ distinct physical qubits, where $d$ is referred to as the code
distance. The ratio $k/n$ is called the encoding rate. If a family of $[[n,k,d]]$ QECCs possess a finite error threshold $p_{\text{th}}$
and errors on physical qubits are independent and occur with probability $p$, then a logical error will occur with probability
$p_L \propto (p/p_{th})^{d/2}$. Therefore as long as $p < p_{th}$, the logical error
rate can be made exponentially small as $d$ is increased. In order to achieve fault-tolerant quantum computation,
it must further be possible to implement an accurate universal set of logical gates on the encoded qubits.

The fault-tolerant storage and processing of quantum information come at significant resource costs in both space overhead,
$n/k$, and the time overhead for implementing logical gates. Families of QECCs that are known to possess a finite error threshold,
code distance $d$ growing with $n$, and constant space overhead, where $n/k$ is a constant independent of code distance $d$, have been proposed
through two basic constructions. The first is in terms of topological codes defined on the cellulation of hyperbolic
space \cite{freedman2002z2, Guth:2014cj}. The second is the hypergraph product code
construction\cite{Tilich2014, Leverrier:2015ju}\footnote{Many other QECCs with constant space overhead
  are also known (see e.g. Ref. \cite{calderbank1996,bravyi2014}), however it is not clear whether they possess a finite error threshold, as
they are not low density parity check (LDPC) codes. The LDPC property guarantees that a stabilizer code
with $d \sim n^\alpha$ for $\alpha > 0$ will have finite error threshold.\cite{kovalev2013}. On the other hand,
many quantum LDPC codes with bounded $d$ in the limit of $n \rightarrow \infty$ are also known.}.

While constant space overhead is known to be possible, it is unclear what the fundamental time overhead must be
for performing a universal set of logical gates. Universal logical gate sets on encoded qubits can in general be
implemented through three known methods: (1) state distillation and gate implementation through
measurements \cite{bravyi2005,nielsen_chuang_2010}, (2) code switching \cite{Paetznick:2013fu}, and
(3) braiding or Dehn twist operations in appropriate classes of topological
codes \cite{Freedman_Larsen_wang_2002,kitaev2003,levin2005,zhwang2010,Koenig:2010do,bonderson2009,barkeshli2016mcg}.

The methods (1) - (3) necessarily require either (a) measurements, which are non-trivial to take fault-tolerantly, together with follow-up
operations that depend on the results of those measurements, or (b) unitary circuits whose depth grows
linearly with the code distance. An exception is the class of protocols recently introduced in Ref. \cite{Zhu:2018CodeLong,zhu2018},
which demonstrate how universal gate sets can be applied through constant depth unitary circuits (with the depth independent
of $n$ and $d$).

To date, it has not been demonstrated how methods (2) or (3) can be combined with QECCs that have constant space
overhead. Moreover, to our knowledge the optimal space-time overhead achieved using method (1) is due to a proposal of
Gottesman \cite{Gottesman:2014ug}, who showed that constant space and time overhead are simultaneously
achievable for \it sequential \rm quantum circuits. This means that a generic quantum circuit of depth $D$, which can be
implemented in $\mathcal{O}(D)$ time using parallel gate operations, could take up to time
$\mathcal{O}(k D)$ if implemented sequentially. Note that for quantum codes with constant encoding rate, $k$ is proportional to $n$ and as such, the circuit depth of the sequential implementation could grow linearly with the system size.

In this paper we consider the most general class of topological codes, which we refer to as Turaev-Viro codes.
  These are based on Turaev-Viro-Barrett-Westbury topological quantum field theories (TQFTs) \cite{turaev1992,barrett1996}
and are stabilized by the Levin-Wen string-net Hamiltonians \cite{levin2005}. This class of TQFTs include, as special cases,
the theories that describe the Kitaev surface code and quantum double models \cite{kitaev2003}.
The use of these TQFTs and the associated Levin-Wen Hamiltonians as quantum codes was
discussed explicitly in Ref.~\cite{Koenig:2010do}.

Here we consider Turaev-Viro codes defined on triangulations of hyperbolic surfaces. As we review below,
the constant negative curvature of the hyperbolic surface allows for a finite encoding rate and thus constant
space overhead for the associated topological code \cite{freedman2002z2}. Such hyperbolic codes can be
implemented in a flat two-dimensional layout of physical qubits by allowing long-range couplings between
the physical qubits -- a possibility allowed by a variety of quantum computing architectures (e.g.~ion traps \cite{Linke:2017bz, Lekitsch:2015ua}, modular architectures of superconducting cavity networks \cite{CampagneIbarcq:2017wq, Kurpiers:2017ub, Axline:2017uq, Chou:2018vz}, Rydberg atoms \cite{Saffman:2010ky, Comparat:2010cb, Maller:2015is, Pichler:2016ec} and silicon photonics \cite{HerreraMarti:2010cu}), and a necessary requirement for any QECC with constant space overhead.  In particular, a recent experimental realization of  hyperbolic circuit QED lattices \cite{cKollar:2018vc} is achieved by utilizing the feature that the cavity quantum bus (used to connect superconducting qubits) can have a wide range of length scales.

The main result of our paper is an explicit protocol for implementing Dehn twist operations in hyperbolic Turaev-Viro codes
through constant depth unitary circuits.  Here constant depth refers to the fact that the depth of the circuit is
independent of $n$ and $d$. We note that throughout this paper, by constant depth unitary circuit, we assume implicitly
that every gate in the circuit also acts on a constant number (independent of $d$ and $n$) of physical qubits. In contrast to the proposal of  Ref.\cite{Gottesman:2014ug} which implements logical gates sequentially via gate teleportation, our approach allows parallel gate operations on encoded qubits via unitary circuits.
Our circuit takes the form of a permutation on qubits, followed by a constant depth circuit which is local with respect to the hyperbolic metric
that abstractly defines the code. The permutation can be implemented with a
depth-two circuit by applying long-range SWAP operations in parallel throughout the code. These results generalize
our earlier work \cite{Zhu:2018CodeLong,zhu2018,Zhu:2017tr} demonstrating that braids and Dehn twists in
general topological codes associated with triangulations of Euclidean space can be implemented by similar constant depth unitary circuits.
The extension to hyperbolic space described here implies that these protocols are also compatible with having
constant space overhead. Our results demonstrate explicitly an advantage in terms of space complexity for implementing the mapping class group
of closed manifolds in topological codes, as compared with the more well-studied braid group of punctures\cite{Beverland:2016bi,nayak2008,kitaev2003}.

Our protocols are inherently protected from errors in the sense that all error strings that are introduced to the system by faulty
physical operations have $\mathcal{O}(1)$ length; moreover, all pre-existing error strings grow by at most
an $\mathcal{O}(1)$ factor. Stated differently, our circuits map local operators to local operators:
an operator with support in a local region $\mathcal{R}$ is mapped to an operator with support in a local
region $\mathcal{R}'$, such that the area of $\mathcal{R}$ and $\mathcal{R}'$ are related by a constant factor
independent of $d$. However, since this constant factor is greater than unity, our circuits may grow error strings
by a constant factor that is larger than unity. In the presence of syndrome measurement errors we thus
require $\mathcal{O}(d)$ rounds of error correction for every $\mathcal{O}(\log d)$ logical gates
that are applied.

Our protocols demonstrate how to apply logical gate operations in hyperbolic codes without
increasing the space overhead. For non-Abelian topological codes, this is the first example of such
protocols. For Abelian topological codes, such as the $\mathbb{Z}_2$ hyperbolic surface code (the extension
of the Kitaev $\mathbb{Z}_2$ toric code to hyperbolic space \cite{breuckmann2016,Breuckmann:2017hy}), a measurement-based approach using
twist defect ancillas was described in Ref.~\cite{lavasani2018}.  The constant depth logical
gates developed here for the general hyperbolic Turaev-Viro code also be applied to the hyperbolic
surface code as a specific case, where a subset of Clifford logical gates \cite{Zhu:2017tr} can be implemented
through constant depth unitary circuits. We note that Ref.~\cite{Breuckmann:2017hy} also discussed
Dehn twists in hyperbolic stabilizer codes, however constant depth protocols that can preserve the constant space
overhead were not presented.

As we discuss, our protocols have important implications for improving the asymptotic scaling of the
space-time overhead required for fault-tolerant universal quantum computation. In particular,
they suggest that universal fault-tolerant quantum computation with constant space overhead
and a time cost of $\mathcal{O}(d D/\log d)$ is possible for a logical quantum circuit of depth $D$.
For certain highly parallel logical quantum circuits, this yields the best asymptotic scaling
known to date.

This paper is organized as follows. In Sec.~\ref{sec:review}, we review the construction of
hyperbolic Turaev-Viro codes. In Sec.~\ref{sec:geogateset}, we sketch the two key
steps for how to implement logical operations corresponding to Dehn twists in hyperbolic Turaev-Viro codes.
The first step requires explicit maps representing Dehn twists on hyperbolic
surfaces, which we present in Sec.~\ref{sec:continuousMaps}. The second step, explained in Sec.~\ref{sec:reTriangulation}
demonstrates how Turaev-Viro codes associated with different triangulations of hyperbolic space can be related to each other
through local constant depth circuits. In Sec.~\ref{sec:FaultTolerance} we discuss the fault-tolerance of these protocols and
and their implications for asymptotic space-time resource costs for universal fault-tolerant quantum computation.
We conclude with a discussion in Sec.~\ref{sec:conclusion}.

\section{Hyperbolic Turaev-Viro code}\label{sec:review}
The Turaev Viro code \cite{levin2005,Koenig:2010do} is a quantum error correcting code which is
defined based on a given unitary fusion category $\mathcal{C}$ and is constructed using qudits that
reside on the edges of a triangulation $\Lambda$ of a surface $\Sigma$. Below we briefly review
the construction of such codes.

Consider a unitary fusion category \cite{zhwang2010} $\mathcal{C}$ with $N$ simple objects. We associate a vector
space $V_{ab}^c$ to any triplet of simple objects $(a,b,c)$, whose dimension corresponds to the fusion rules:
\begin{equation}
  N_{ab}^c=\text{dim } V_{ab}^c.
\end{equation}
We may write fusion rules formally as:
\begin{equation}
  a\times b=\sum_c N_{ab}^c~c.
\end{equation}
Associativity of the fusion rules gives a constraint on the fusion coefficients $N_{ab}^c$.
In particular, the vector spaces $\bigoplus_{e} V^e_{ab}\otimes V^{d}_{ec}$ and $\bigoplus_{f} V^f_{bc}\otimes V^{d}_{af}$ are isomorphic.
The unitary map between these two vector spaces are the $F$-symbols of the theory,
\begin{equation}
  F^{abc}_d: \bigoplus_{e} V^e_{ab}\otimes V^{d}_{ec}\mapsto \bigoplus_{f} V^f_{bc}\otimes V^{d}_{af}
\end{equation}

\begin{figure}
  \includegraphics[width=0.9\columnwidth]{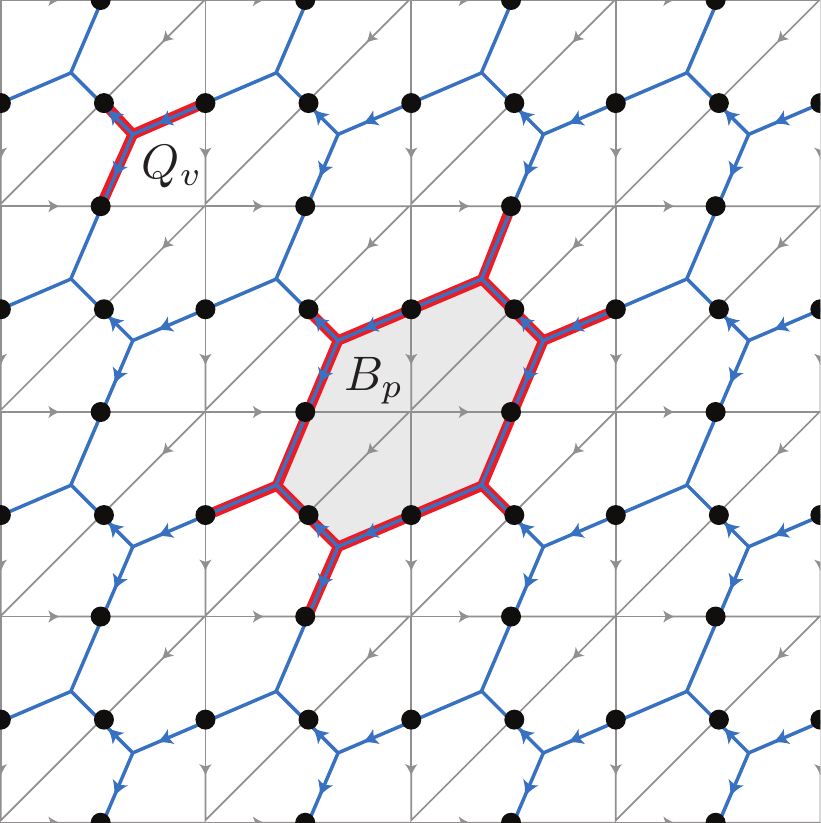}
  \caption{A triangulation $\Lambda$ drawn by light gray solid lines
    and its dual trivalent graph $\hat \Lambda$ drawn by dark blue
    lines. The arrows on the edges define the branching structure.
    Black dots represent the physical qubits. Examples of vertex operators $Q_v$
    and flux operators $B_p$ and their support are also illustrated.}
  \label{fig:tvc}
\end{figure}

Consider a surface $\Sigma$ together with a triangulation $\Lambda$ (see Fig. \ref{fig:tvc}) of $\Sigma$.
The triangulation is also equipped with a branching structure (i.e. a local ordering of vertices).
To each edge of the triangulation we assign an $N$-state qudit with states labeled
$\ket{a_i}$, where $a_i$ are the simple objects in $\mathcal{C}$.
For simplicity of the construction we assume that the fusion rules $N_{ab}^c$ are $0$ or $1$, although
this can be easily generalized by including additional degrees of freedom at the vertices.

To define the code, it is simpler to work with the  cellulation dual to $\Lambda$, which we denote by
$\hat \Lambda$. Note that since $\Lambda$ is a triangulation of the surface, $\hat \Lambda$ is
a trivalent graph. If a qudit is in state $\ket{a}$, we say a type $a$ string is passing through the
corresponding edge on $\hat \Lambda$ and label the edge by $a$. The wave functions of the code
space then can be seen as superpositions of string-net configurations that are
consistent with certain string branching rules\cite{levin2005}.

The code space $\mathcal{H}_{\Lambda}(\Sigma)$ is a subspace of the full Hilbert space of the physical qudits.
The topological nature of the code guarantees that different choices of triangulations $\Lambda$ and branching
structures yield isomorphic code subspaces.

In particular $\mathcal{H}_\Lambda(\Sigma)$ corresponds to the ground state subspace of a local
Levin-Wen Hamiltonian \cite{levin2005}:
\begin{equation}\label{eq:levwen}
  H_{\hat \Lambda}= - \sum_v Q_v -\sum_p B_p,
\end{equation}
where $\sum_v$ and $\sum_p$ sum over all vertices and plaquettes of $\hat \Lambda$ respectively (see Fig. \ref{fig:tvc}).
All $Q_v$ and $B_p$ operators commute with each other. One can think of this model as a generalization from
the abelian surface code to arbitrary (abelian and non-abelian) non-chiral topological orders in 2D. We note that
the Levin-Wen Hamiltonian as defined in Ref.~\cite{levin2005} requires a certain tetrahedral symmetry for the
$F$-symbols, which makes the branching structure of the triangulation unnecessary, although the construction
can be generalized to relax this condition.

The vertex operator $Q_v$ is a local projection operator which ensures that the ground state wave function is
consistent with the branching rules of the theory. The action of $Q_v$ on a state $\ket{\psi}$ depends only on
the states of qudits that reside on the edges which are incident to $v$ and is defined as:
\begin{align}\label{branching_rules}
Q_v
\begin{tikzpicture}[baseline={([yshift=-.5ex]current  bounding  box.center)}]
\draw[thick]  (-0.8, -0.5) -- (-0.8, 0.5);
\draw[dualblue,thick,middlearrow={stealth reversed}] (0,0) --  (0.8,0);
\draw[dualblue,thick,middlearrow={stealth reversed}]  (0,0) -- (-0.3,0.5)   ;
\draw[dualblue,thick,middlearrow={stealth reversed}] (-0.3,-0.5) -- (0,0)   ;
\draw (0.5,0.15) node {$a$};
\draw (-0.45,0.45) node {$b$};
\draw (-0.45,-0.45) node {$c$};
\draw (0.1,-0.2) node {$v$};
\draw[thick]   (1, -0.5) -- (1.2, 0);
\draw[thick]   (1,  0.5) -- (1.2, 0);
\end{tikzpicture}
= N_{ab}^c
\begin{tikzpicture}[baseline={([yshift=-.5ex]current  bounding  box.center)}]
\draw[thick]  (-0.8, -0.5) -- (-0.8, 0.5);
\draw[dualblue,thick,middlearrow={stealth reversed}] (0,0) --  (0.8,0);
\draw[dualblue,thick,middlearrow={stealth reversed}]  (0,0) -- (-0.3,0.5)   ;
\draw[dualblue,thick,middlearrow={stealth reversed}] (-0.3,-0.5) -- (0,0)   ;
\draw (0.5,0.15) node {$a$};
\draw (-0.45,0.45) node {$b$};
\draw (-0.45,-0.45) node {$c$};
\draw (0.1,-0.2) node {$v$};
\draw[thick]   (1, -0.5) -- (1.2, 0);
\draw[thick]   (1,  0.5) -- (1.2, 0);
\end{tikzpicture}
\end{align}
The action of the flux operator $B_p$ on a wave function depends only on the edges which are incident on the vertices of $p$ and is defined using the $F$-symbols. It can be thought of as the generalization of the local plaquette operators in $\mathbb{Z}_2$ surface codes. Its exact form is rather complicated and we refer the interested reader to References \cite{levin2005,Koenig:2010do} for a complete and through review of the Turaev-Viro codes.

Just like the surface code, the eigenstates of the Hamiltonian \eqref{eq:levwen} can be realized
in an active error correction approach by repeated measurement of all $Q_v$ and $B_p$ operators,
each of which can be performed by a local constant depth circuit together with an ancilla.
 The measurement results are then used to detect and correct possible errors. Designing explicit
 quantum circuits for syndrome measurements and finding fault tolerant error correction schemes
 for specific variants of Turaev-Viro codes are subjects of ongoing research \cite{Bonesteel:2012fl,feng2015non,Burton:2017gr}.

A hyperbolic Turaev-Viro code is a Turaev-Viro code which is defined on a triangulation of a closed hyperbolic
surface $\Sigma$. A closed hyperbolic surface is a closed surface endowed with a Riemannian metric of constant curvature $-1$.
Due to the Gauss-Bonnet theorem, the area of such a surface can be found from its genus $g$:
\begin{equation}\label{eq:gauss}
  A_\Sigma=4\pi(g-1).
\end{equation}
The number of logical qubits (i.e.~$\log [\text{dim } \mathcal{H}_\Lambda(\Sigma)]$)
for a topological quantum code on a closed genus $g$ surface is proportional to $g$.
On the other hand, for a fine triangulation with bounded geometry, where by bounded geometry
we mean that the edge lengths and angles are bounded from above and below, the number of
physical qubits $n$ is proportional to the surface area $A_\Sigma$ and the code distance $d$ scales
as $\log(n)$ \cite{freedman2002z2}. Therefore,  according to \eqref{eq:gauss}, by using hyperbolic surfaces of increasing
genus we can construct a family of hyperbolic Turaev-Viro codes with constant encoding rate and increasing code distance. Note that the
encoding rate for a quantum code defined on a Euclidean surface is $\mathcal{O}(1/d^2)$ and
goes to $0$ as one goes to large distances \cite{bravyi2010tradeoffs}.

We further note that the existence of a lower bound on the angles in the fine triangulation also ensures that the associated
error-correcting code is in essence a low-density parity check (LDPC) code, which has a low-weight (upper-bounded)
plaquette syndrome $B_p$ on the dual trivalent graph $\hat{\Lambda}$ (the vertex operator $Q_v$ is always weight-3
due to the definition of a triangulation or equivalently the trivalent structure of its dual graph). As is the case with
stabilizer codes \cite{kovalev2013}, the LDPC property is important for the possibility of an error threshold in the
presence of noise during the syndrome measurements. A particular type of triangulation satisfying this
property is the Delaunay triangulation, which is a unique triangulation for a given set of vertices that
maximizes the smallest angle among all possible triangulations \cite{Hjelle:2006wr}.   The hyperbolic
Delaunay triangulation can be constructed numerically starting from a randomly picked distribution of
vertices with a given average density in hyperbolic space.   Then standard classical computer algorithms
generating a Delaunay triangulation, such as the radial-sweep algorithm, can be applied \cite{Hjelle:2006wr}.
Some detailed studies of hyperbolic Delaunay triangulations can also be found in
Ref.~\cite{Hyperbolic_Delanunay_2013,
  Hyperbolic_Delanunay_closed_2013}.

We note that numerical evidence for a finite error threshold in hyperbolic surface (stabilizer) codes was established
in Ref. \cite{breuckmann2016,Breuckmann:2017hy}.

\section{Geometric gate sets for hyperbolic Turaev-Viro codes}\label{sec:geogateset}

Consider a Turaev-Viro code defined using a unitary fusion category $\mathcal{C}$ on a triangulation
$\Lambda$ of a closed surface $\Sigma$ with genus $g$. It is well-known that the code space
$\mathcal{H}_\Lambda(\Sigma)$ forms a non-trivial representation of the mapping class group (MCG) of the surface
$\Sigma$. In other words, elements of the MCG implement certain non-trivial operations on the code space.
Recall that the MCG is the group of homeomorphisms of the surface modulo those homeomorphisms
that are continuously connected to the identity \footnote{The MCG can also be defined using diffeomorphisms;
both definitions are equivalent \cite{farb2011primer}.}. We call the set of such operations the \textit{geometric gate set}.
Gates corresponding to MCG operations are naturally topologically protected
(and thus can be made fault-tolerant) and can be implemented through a variety of
methods \cite{nayak2008,zhwang2010,Koenig:2010do,barkeshli2016mcg,Zhu:2017tr,Zhu:2018CodeLong}.
For certain codes, such as the Fibonacci Turaev-Viro code, the geometric gate set forms a universal gate set \cite{Freedman_Larsen_wang_2002}.

Here we consider a way of implementing MCG elements in terms of constant depth unitary circuits
that is closely related to the method proposed in \cite{Zhu:2018CodeLong}, although our presentation
below is somewhat different and more general. This method can then be applied to the case of hyperbolic codes which
yield constant space overhead.

Let $U$ be a mapping class group element of the surface $\Sigma$. We denote its representation on
the code space $\mathcal{H}_\Lambda(\Sigma)$ by $\mathcal{U}$. For a given $U$, one can implement
$\mathcal{U}$ using the following procedure:

\begin{itemize}
  \item (Step 1) Let $f_U:\Sigma\to \Sigma$ be a specific homeomorphism representing $U$. We move the vertices of $\Lambda$ using $f_U$, and connect them as they were connected originally to get a new triangulation of $\Sigma$ which we denote by $\Lambda'$. This operation corresponds to a permutation of the physical qubits. If the qubits are mobile, this transformation can be carried out by shuttling the qubits around. Otherwise, it can be implemented as a depth-two circuit by using long-range SWAPs in parallel throughout the system \cite{Zhu:2018CodeLong}.

  \item (Step 2) Since the Turaev-Viro code was defined using the triangulation $\Lambda$, after the first step the wave function of the system would no longer be in the original code space $\mathcal{H}_\Lambda(\Sigma)$. Rather it would be associated to the code space $\mathcal{H}_{\Lambda'}(\Sigma)$ of a different triangulation $\Lambda'$.
    To remedy this, we apply a local quantum
  circuit that effectively implements a local geometry deformation and transforms the code defined on the
  $\Lambda'$ triangulation back to the one defined on the original $\Lambda$ triangulation. We will show in
  subsequent sections that this retriangulation can be performed via a \it constant depth \rm local quantum circuit.
   If we regard this transformation as a homeomorphism of the surface $\Sigma$, it would be equivalent
   to the trivial element of MCG, which ensures that it will not result in another nontrivial transformation
   on top of the map $U$. Details of the geometry deformation circuit are explained in Section \ref{sec:reTriangulation}.
\end{itemize}

As an example, take $\Sigma$ to be the torus $T^2$ with a regular triangulation $\Lambda$ which is used to define the
Turaev-Viro code. To construct the torus, we can take a square of side $1$ and identify the opposite sides.

\begin{figure}[t]
  \includegraphics[width=\columnwidth]{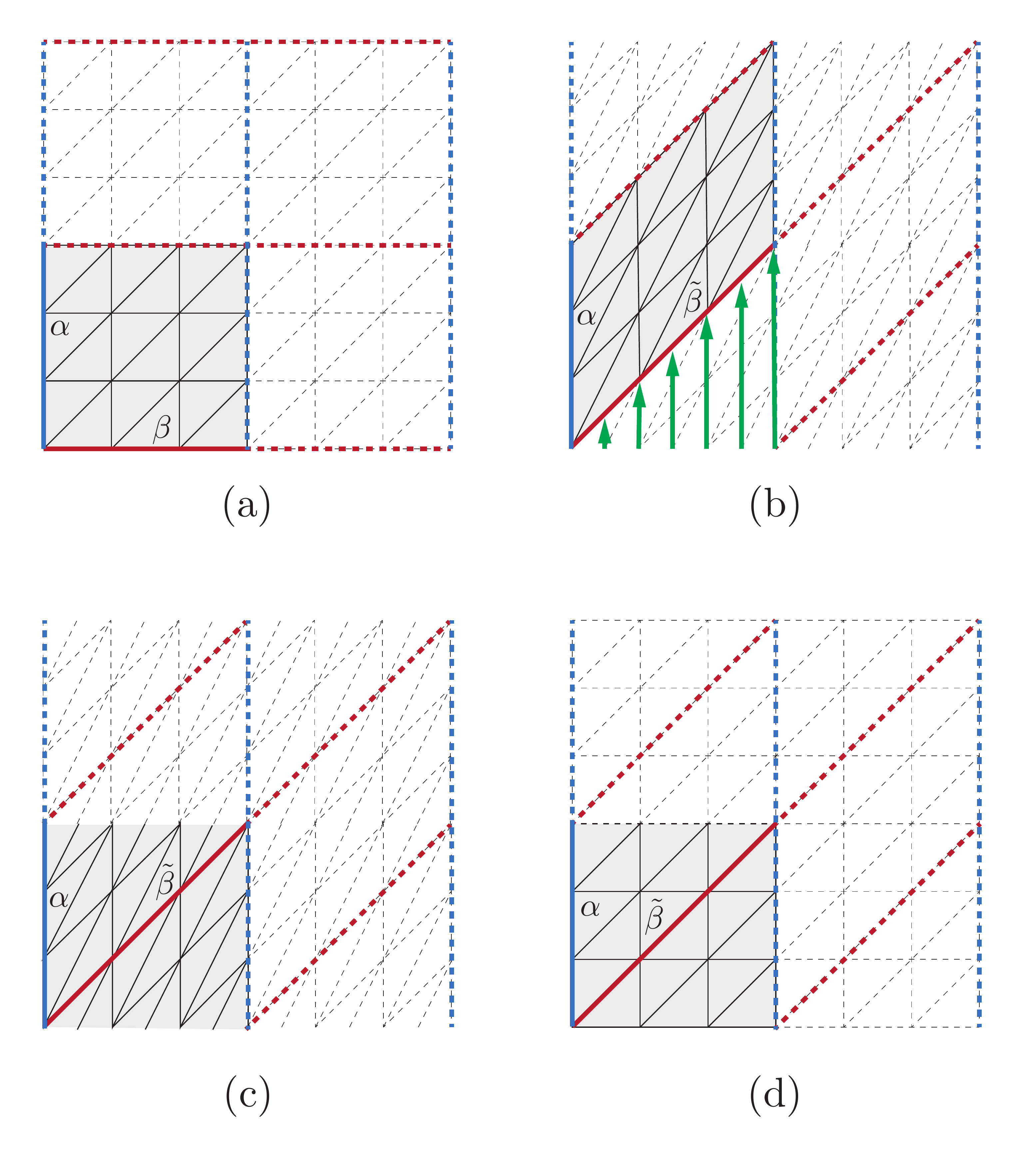}
  \caption{(a) The complex plane $\mathbb{C}$ as the universal cover of the torus $T^2$. Blue and red lines correspond to the canonical loops on the torus. The triangulation $\Lambda$ of the torus is shown as well. The shaded region can be taken as the fundamental domain of the covering. By applying $f_U(x+iy)=x+i(x+y)$ (green arrows), the shaded region in (a) maps to the sheared region in (b) and the $\beta$ loop maps to $\tilde \beta$ while the $\alpha$ loop remains intact. By looking at the original fundamental domain as in (c) it becomes clear that $\tilde \beta$ goes around both handles. Application of the local geometry deformation circuit then recovers the original triangulation as shown in (d) and maps the wave function back into the original code space.}
  \label{fig:lattice_torus}
\end{figure}

Alternatively, we can start with the complex plane $\mathbb{C}$ and identify points according
to equivalence relations $z \sim z+1$ and $z\sim z+i$. We can use these identification
rules to define a universal covering map from $\mathbb{C}$ to $T^2$. A triangulation of
$T^2$ then translates to a triangulation of the complex plane (see Fig. \ref{fig:lattice_torus}a).

Let $U=D_\alpha$ be the Dehn twist along $\alpha$, the meridional loop of the torus(for a brief review of Dehn twists, see Appendix \ref{sec:dehn}). Consider the shearing
map $f_D(x+iy)=x+i(x+y)$. It is easy to verify that this map respects the equivalence relations and corresponds to a Dehn twist along
the $\alpha$ loop. If we move (permute) the qubits according to $f_D$, we get the configuration shown in Fig. \ref{fig:lattice_torus}b.
Note that as a result of this map, the string along the $\beta$ loop now encircles both handles while the string along the $\alpha$ loop remains unchanged, as one would expect form a Dehn twist along $\alpha$. As a result of the previous step, the triangulation of the torus has been changed, as one can see by comparing Fig.~\ref{fig:lattice_torus}(c) and Fig.~\ref{fig:lattice_torus}(a). To compensate for that, we will apply a local unitary circuit, which corresponds to the trivial element of the MCG, to restore the original triangulation without applying any further logical gate. The final result is shown in Fig.~\ref{fig:lattice_torus}(d).

In Ref \cite{zhu2018, Zhu:2018CodeLong}, the above procedure has been used extensively to implement MCG elements by finite depth quantum circuits in QECCs which are defined on Euclidean surfaces. The main result of this work is that the same basic idea can be used to implement logical gates in QECCs which are defined on a hyperbolic surface. In the following we are going to explain in detail how one can implement geometric gates in hyperbolic Turaev-Viro codes.

Let $\Sigma$ denote a hyperbolic surface that is used to define the hyperbolic Turaev-Viro code. Since the MCG can be generated by the Dehn twists around the handles of $\Sigma$, to implement an arbitrary geometric gate it suffices to be able to implement
Dehn twists around the handles of $\Sigma$ \cite{Zhu:2018CodeLong}.

In Section \ref{sec:continuousMaps}, we construct specific diffeomorphisms that correspond to the basic Dehn twists, which then can be used to carry out Step 1 of the above procedure. Next, in Section \ref{sec:reTriangulation}, we introduce the local finite depth quantum circuit that converts two given triangulations to one another. By combining the results of these two sections and following the above procedure, one can implement the representation of any basic Dehn twists on the code space $\mathcal{H}_\Lambda(\Sigma)$, and hence implement any geometric gate by a constant depth unitary circuit.

\section{Continuous Maps for Dehn Twists }\label{sec:continuousMaps}

First we concentrate on the $g=2$ case. After developing the maps for the simplest case, we show how these maps can be generalized for a surface of arbitrary genus.
\subsection{Dehn Twists on a Double torus}

\begin{figure*}[t]
  \includegraphics[width=2\columnwidth]{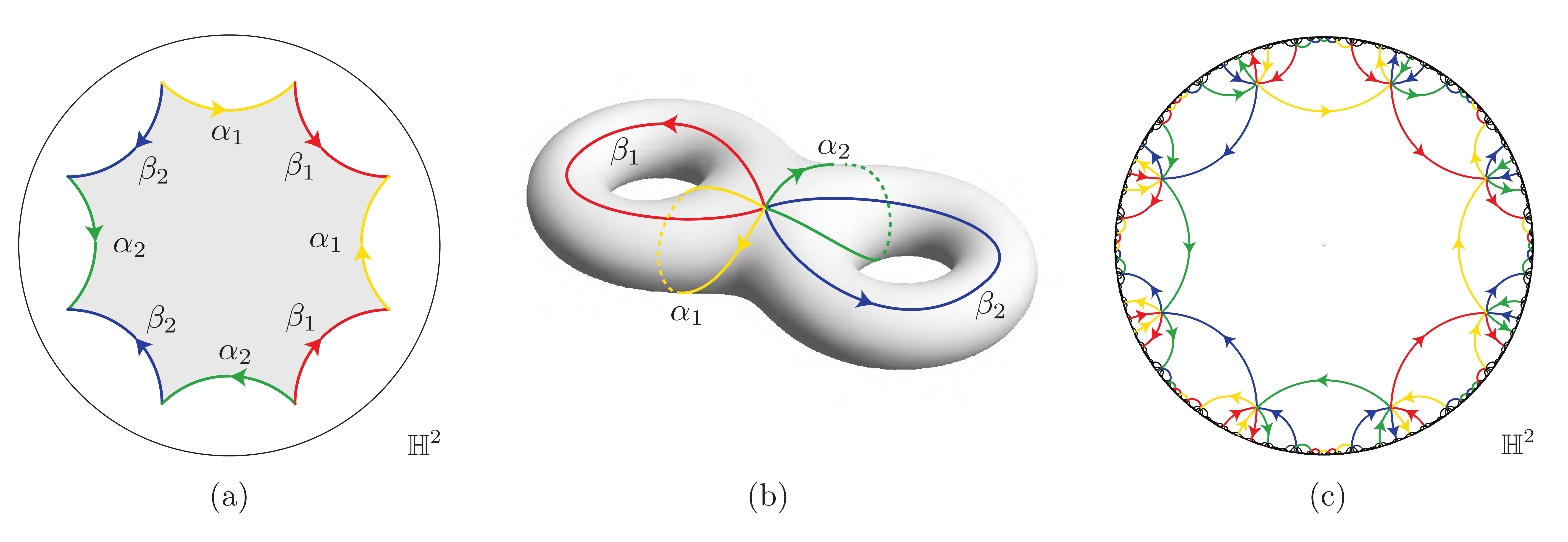}
  \caption{(a) The canonical octagon plotted on the Poincare disk model of the hyperbolic plane. Edges with the same
    color (label) have to be identified such that the arrows on the two edges line up to each other.
    (b) The $\Sigma_2$ surface resulting from identifying the edges of the canonical octagon. (c) Tilling of the hyperbolic plane by the canonical octagon.}
  \label{fig:canonical}
\end{figure*}

Let $\tilde \Sigma_2$ be an arbitrary genus $2$ surface. As a surface with negative Euler characteristic,
$\chi(\tilde\Sigma_2)=-2$, it admits a hyperbolic metric, i.e. a complete finite area Riemannian metric of
constant negative curvature $-1$.

One way to define a hyperbolic metric on $\tilde \Sigma_2$ is to start with the regular hyperbolic octagon on
$\mathbb{H}^2$ whose interior angles sum to $2\pi$, known as the Fricke canonical polygon \cite{fricke1897vorlesungen,keen1966canonical}. If we identify every other edge as shown in Fig. \ref{fig:canonical}a with the arrows specifying how the edges should be lined up, we obtain a genus $g=2$ hyperbolic surface $\Sigma_2$
shown in Fig. \ref{fig:canonical}b, which is homeomorphic to $\tilde \Sigma_2$. According to Ref. \cite{munkres1960obstructions}, the homeomorphism can be upgraded to a diffeomorphism and thus induces a hyperbolic metric on $\tilde{\Sigma}_2$. From now on, we concentrate on $\Sigma_2$, knowing that our statements about $\Sigma_2$ can be generalized to the $\tilde\Sigma_2$ surface using the aforementioned diffeomorphism.

If we instead consider moving the vertices of the octagon to more general locations on the
hyperbolic plane to obtain an irregular octagon, we move through Teichm\"uller space, which is the space of hyperbolic
metrics of the closed hyperbolic surface.

As a result of the identification scheme, all vertices of the canonical polygon represent a single point on $\Sigma_2$ and thus the sides of the polygon correspond to closed loops with a common base point on $\Sigma_2$ (see Fig. \ref{fig:canonical}b). Conversely, if we start with the double torus in Fig. \ref{fig:canonical}b and cut the surface along these loops, we will obtain the octagon in Fig. \ref{fig:canonical}a. In fact, these loops can be taken as the generators of the fundamental group $\pi_1(\Sigma_2)$,
\begin{equation}\label{eq:fg}
  \pi_1(\Sigma_2)=\langle \alpha_1,\beta_1,\alpha_2,\beta_2\,|\,
 \prod_{i=1}^2 \alpha_i \beta_i \alpha_i^{-1}\beta_i^{-1}=1 \rangle.
\end{equation}

An element of the MCG will take these loops to some other loops on $\Sigma_2$, and thus naturally induces a map over $\pi_1(\Sigma_2)$. More precisely, it can be shown \cite{dehn2012papers,nielsen1927untersuchungen} that the mapping class group is isomorphic to the group of outer automorphisms of the fundamental group,
\begin{equation}\label{eq:DNB}
  \text{MCG}(\Sigma_2)\approx \text{Out}(\pi_1(\Sigma_2))=\text{Aut}(\pi_1(\Sigma_2))/\text{Inn}(\pi_1(\Sigma_2)),
\end{equation}
where $\text{Aut}(G)$ and $\text{Inn}(G)$ denote the automorphism group and inner automorphism group of $G$ respectively.
The equivalence under $\text{Inn}(\pi_1(\Sigma_2))$ is related to the fact that in general, homeomorphisms of
$\Sigma_2$ will not keep the base point of the loops in $\pi_1(\Sigma_2)$ fixed.

Let $U$ denote an arbitrary element of the mapping class group and hence an equivalence class of homeomorphisms on $\Sigma_2$. Due to Eq. \eqref{eq:DNB}, $U$ also corresponds to an equivalence class of automorphisms of $\pi_1(\Sigma_2)$. In the rest of this paper, in an abuse of notation we use the same symbol $U$ to denote both the equivalence classes of
homeomorphisms and $\text{Out}(\pi_1(\Sigma_2))$ and also representatives of these classes.

Consider the canonical octagon on the hyperbolic plane. By attaching a copy of the octagon on each edge according to the identification rules and continuing this procedure indefinitely for the edges of the newly added octagons, one will end up with the $\{ 8,8\}$ tiling of the hyperbolic plane (see Fig. \ref{fig:canonical}c). The result can be used to define a covering map $p:\mathbb{H}^2\longrightarrow \Sigma_2$ which along with $\mathbb{H}^2$ makes up the universal cover of $\Sigma_2$.

In what follows we provide explicit expressions for the homeomorphisms of $\Sigma_2$ to itself corresponding to the Dehn twists along the primary loops of $\Sigma_2$.

\subsubsection{Dehn twists along the $\alpha$ and $\beta$ loops}\label{sec:Da}

We start with the Dehn twist along $\alpha_1$.
\begin{equation}
  D_{\alpha_1}:\Sigma_2 \longrightarrow \Sigma_2
\end{equation}
To find an element of $\text{Aut}(\pi_1(\Sigma_2))$ which represents $D_{\alpha_1}$, it is enough to see how it acts on the canonical loops of the $\Sigma_2$ surface. To this end we can use the Dehn surgery method described in Appendix \ref{sec:dehn}. However, for simplicity, first we push the $\alpha_1$ slightly to the left to detach it from the $\alpha_2$ and $\beta_2$ loops and consider the twist map along this new loop. Note that the Dehn twist $D_{\alpha_1}$ depends only on the isotopy class of $\alpha_1$. Then, as one can verify by looking at Fig. \ref{fig:canonical}b and using the Dehn surgery method, $D_{\alpha_1}$ maps $\beta_1$ to $\beta_1 \alpha_1^{-1}$ and leaves all the other canonical loops invariant. Note that we use the left to right convention for loop multiplication; if $f$ and $g$ are two loops with a common base point, $fg$ corresponds to a loop that traces $f$ first and then $g$.

Instead of specifying $D_{\alpha_1}$, we provide an explicit form for its lift to the covering space $D^\ast_{\alpha_1}:\mathbb{H}^2\longrightarrow \mathbb{H}^2$ such that the diagram below commutes:
\begin{equation}\label{eq:dehndiag}
  \begin{tikzcd}
  \mathbb{H}^2 \arrow[r,"D^\ast_{\alpha_1}"] \arrow[d,"p"]
  & \mathbb{H}^2 \arrow[d, "p"] \\
  \Sigma_2 \arrow[r, "D_{\alpha_1}"]
  & \Sigma_2
  \end{tikzcd}
\end{equation}
To this end, we define how $D^\ast_{\alpha_1}$ acts on the points of a fundamental domain. Its action on the other points of $\mathbb{H}^2$ are defined accordingly to ensure  commutativity of the diagram in Eq.\eqref{eq:dehndiag}.

\begin{figure*}[t]
  \includegraphics[width=2\columnwidth]{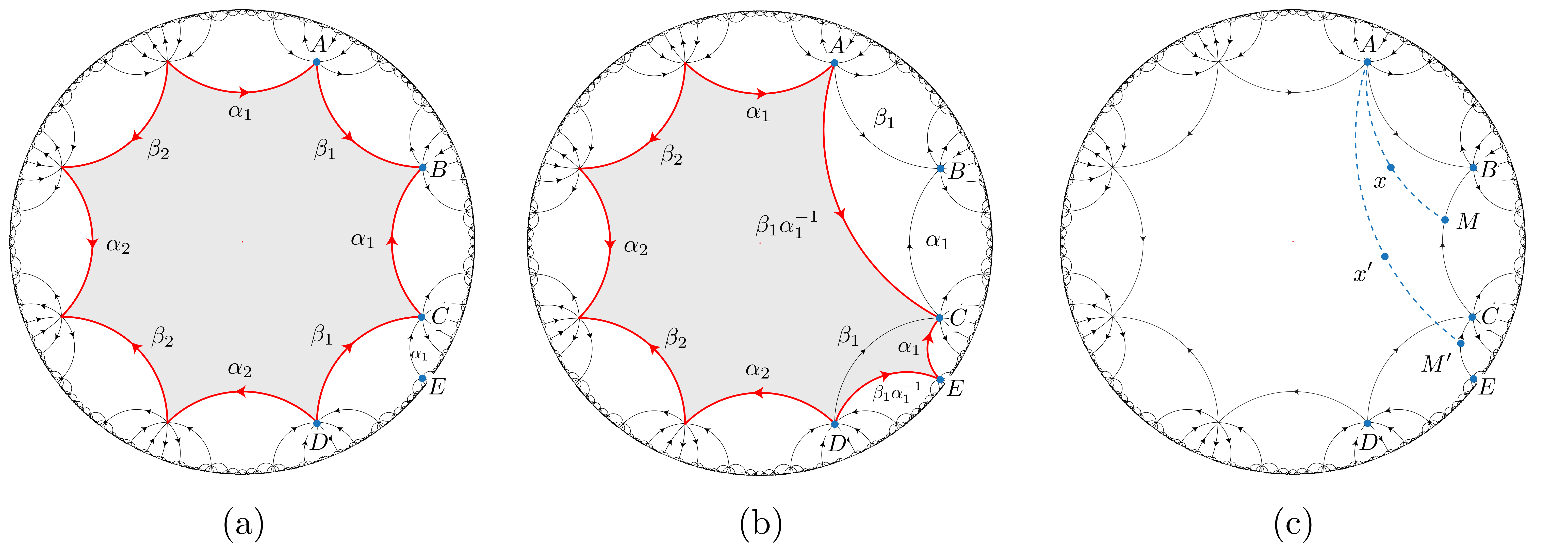}
  \caption{(a) The fundamental domain (the shaded region) can be identified with $\Sigma_2$. We define the action of the shearing map on the points in this region. The action of the map on the other points is defined according to the identification scheme. (b) Image of the fundamental domain under the shearing map $D^\ast_{\alpha_1}$. (c) Example of the action of the map: the point $x$ inside the triangle $\Delta ABC$ maps to the point $x'$ inside $\Delta ACE$. See the text for details.}
  \label{fig:dalpha}
\end{figure*}

We take the fundamental domain to be the canonical octagon (shaded region in Fig. \ref{fig:dalpha}a). The action of $D^\ast_{\alpha_1} $ on the fundamental domain gives the sheared octagon shown in Fig. \ref{fig:dalpha}b. As one can easily verify, it transforms the $\beta_1$ loop  to $\beta_1 \alpha_1^{-1}$ as desired, while all the other loops remain unchanged. In particular note that the $\alpha_1$ loop is mapped to itself and thus has not changed.

We define the map $D^\ast_{\alpha_1} $ more precisely as follows. We map $\Delta ABC$ in Fig. \ref{fig:dalpha}b to $\Delta ACE$ and $\Delta ACD$ to $\Delta AED$, where $\Delta ABC$ denotes the hyperbolic triangle made by connecting  $A$,$B$ and $C$ via geodesics on $\mathbb{H}^2$. All the other regions in the octagon are left untouched. Mapping $\Delta ABC$ to $\Delta ACE$ is done as follows. Consider an arbitrary point, $x$, inside $\Delta ABC$ (see Fig. \ref{fig:dalpha}c). To find its image $x'=D^\ast_{\alpha_1}(x)$, first draw the geodesic line that passes through $A$ and $x$, and continue it to find its crossing point $M$ with the line $BC$. Now choose $M'$ on line $CE$ such that $|CM'|/|CE|=|BM|/|BC|$. By $|PQ|$ we mean the length of the geodesic line connecting $P$ and $Q$, measured using the hyperbolic metric. Finally, we choose $x'$ on the $AM'$ line such that $|Ax'|/|AM'|=|Ax|/|AM|$.
 $\Delta ACD$ is mapped similarly to $\Delta AED$.

 It is clear that this map is continuous for the points inside the fundamental domain. It is also straightforward to
 check that this map is consistent with edge identification rules and thus is continuous throughout the $\mathbb{H}^2$
 plane. Moreover, the diagram in Eq.\eqref{eq:dehndiag} commutes by construction and hence $D_{\alpha_1}$ is
 continuous on $\Sigma_2$. Furthermore, as we explicitly verify in Appendix \ref{sec:areacalc}, this map does not
 change the area of any region by more than a constant factor. This is an important property which allows the second step of our Dehn twist
 protocol, as we explain in subsequent sections.

Dehn twists along $\beta_1$, $\alpha_2$ and $\beta_2$ are defined in a similar manner.

\subsubsection{Dehn twists along the $\gamma$ loop}

\begin{figure*}[t]
  \includegraphics[width=2\columnwidth]{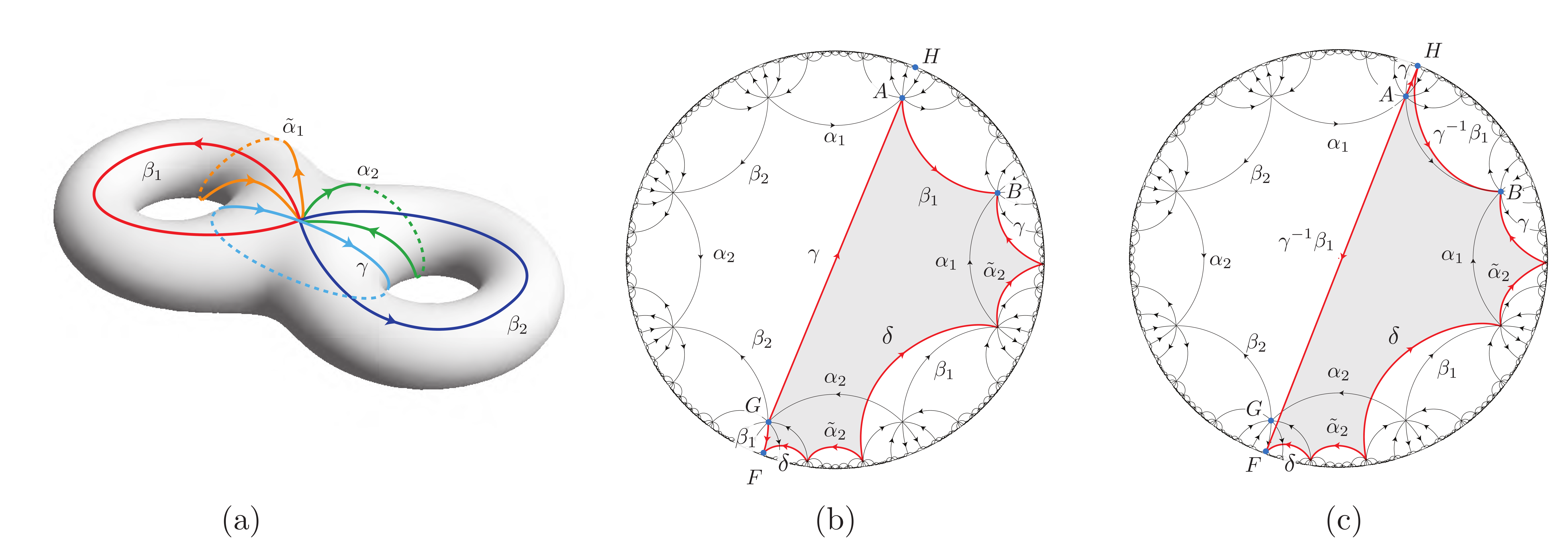}
  \caption{(a) $\tilde\alpha_1=\beta_1 \alpha_1 \beta_1^{-1}$ and $\gamma$ loops on the $\Sigma_2$ surface. (b) The shaded region can be taken as the fundamental domain of the covering map. (c) Image of the the shaded region in (b) under the shearing map $D^\ast_\gamma$.}
  \label{fig:Dg}
\end{figure*}

To generate all elements of the MCG we need the Dehn twist along the $\gamma$ loop as well (see Fig. \ref{fig:Dg}a). The canonical octagon which we used to define $D_{\alpha_1}$ has two important features: first, $D_{\alpha_1}$ only changes the $\beta_1$ sides while leaving other sides of the octagon invariant; second, $\alpha_1$ and $\beta_1$ were neighboring sides of the octagon. However, since both $\beta_1$ and $\beta_2$ transform non-trivially under $D_\gamma$ and since the $\gamma$ loop is not one of the polygon's sides, the action of $D_\gamma$ on the canonical octagon is not as simple as $D_{\alpha_1}$ and looks rather complicated. Therefore to construct $D_\gamma$, it is easier to work with a different fundamental domain. To find the appropriate fundamental domain, we cut $\Sigma_2$ along a new set of loops rather than the standard $\alpha$'s and $\beta$'s.

Let $\tilde{\alpha_i}$ denote the $\alpha_i$ loop translated through $\beta_i$:
\begin{equation}\label{eq:tilde}
  \tilde{\alpha_i}=\beta_i\alpha_i\beta_i^{-1}.
\end{equation}
$\tilde{\alpha}_1$ is illustrated in Fig. \ref{fig:Dg}a. Note that we can write $\gamma$ as:
\begin{equation}
\gamma=\alpha_2^{-1} \tilde{\alpha}_1=\alpha_2^{-1}\beta_1 \alpha_1\beta^{-1}_1.
\end{equation}
We also define $\delta$ as:
\begin{equation}
  \delta=\beta_2\beta_1,
\end{equation}
which represents a loop that encircles both holes of $\Sigma_2$.

As was the case for the $\alpha_1$ Dehn twist, to find an automorphism of $\pi_1(\Sigma_2)$ corresponding to $D_\gamma$ first we push the $\gamma$ loop shown in Fig. \ref{fig:Dg}a slightly to the right and then use the Dehn surgery method to find its action on various loops. We remark that if we used another loop, e.g. if we pushed the $\gamma$ loop slightly to the left instead, we would find the same automorphism up to an action of $\text{Inn}(\pi_1)$. Note that due to Eq. \eqref{eq:DNB}, all such maps represent the same element of the mapping class group.

The representative automorphism induced by $D_\gamma$ on $\pi_1(\Sigma_2)$ can then be summarized in the following four equations:
\begin{alignat}{2}\label{eq:Dg}
  &D_\gamma(\alpha_1)= \alpha_1,\quad D_\gamma(\beta_1)=\gamma^{-1}\beta_1,\nonumber \\
  &D_\gamma(\tilde \alpha_2)=\tilde \alpha_2, \quad D_\gamma(\beta_2)=\beta_2 \gamma.
\end{alignat}
Note that $D_\gamma$ leaves the $\delta$ loop invariant. To find the appropriate fundamental domain, we trade the $\{\alpha_1,\beta_1,\alpha_2, \beta_2\}$
loops with $\{\gamma,\beta_1,\tilde \alpha_2,\delta\}$. The group relation in Eq.\eqref{eq:fg} can be expressed in terms of these loops as well:
\begin{equation}\label{eq:modified_rel}
  \tilde{\alpha}_2^{-1}\delta^{-1}\tilde{\alpha}_2\delta\beta_1^{-1}\gamma \beta_1\gamma^{-1}=1.
\end{equation}
Note that $\{\gamma_1,\beta_1\}$ as well as $\{\tilde\alpha_2,\delta\}$ have algebraic intersection $1$ while the two sets
are mutually detached, i.e. the algebraic intersection number of a loop from the first set and a loop from the second set is $0$.
So the $\{\gamma,\beta_1,\tilde \alpha_2,\delta\}$ loops could have been taken as the primary loops of $\Sigma_2$ in the first place.

This in turn suggests using an octagon with its sides following the $\delta$,$\tilde{\alpha}_2$,$\beta_1$ and $\gamma$ loops.
Such an irregular octagon is shown in Fig. \ref{fig:Dg}b. As one can easily verify, this can be taken as the fundamental domain
of the mapping $p$. Moreover, it has the features we are looking for: under the action of $D_\gamma$, only the $\beta_1$
loop gets deformed and, furthermore, the $\gamma$ and $\beta_1$ loops are represented by neighboring sides of the polygon.

$D^\ast_\gamma$ (the lift of $D_\gamma$ to $\mathbb{H}^2$) shears the octagon shown in Fig.\ref{fig:Dg}b to the one shown
in Fig.\ref{fig:Dg}c. More precisely, $\Delta FGA$ is mapped to $\Delta FAH$ and $\Delta FAB$ is deformed to $\Delta FHB$.
Mapping the triangles is done through the same procedure described in Section \ref{sec:Da}. The action of $D^\ast_\gamma$
on the other points of $\mathbb{H}^2$ is then defined according to the identification rules.
It is straightforward to verify continuity of the map. Moreover, as we verify in Appendix \ref{sec:areacalc},
these maps do not change the area of any region of the surface by more than a constant factor.

Any element of the MCG of the double torus can be generated using $D_{\alpha_i}$, $D_{\beta_i}$ and $D_\gamma$.
In the next section we discuss how these constructions generalize to higher genus surfaces.

\subsection{Dehn Twists on Genus g surface}

A hyperbolic genus $g$ surface can be obtained by identifying every other edge of a $4g$-gon, whose angles sum to $2\pi$, in hyperbolic space.
The space of different hyperbolic metrics, Teichm\"uller space, corresponds to inequivalent choices of the locations
of the vertices of the $4g$-gon \cite{farb2011primer}. Here we consider the canonical $4g$-gon, i.e. a regular $4g$-gon on $\mathbb{H}^2$.
The sides of the polygon can be used to generate the fundamental group of the $\Sigma_g$ surface:
\begin{equation}
  \pi_1(\Sigma_g)=\langle \alpha_1,\beta_1,\cdots,\alpha_g,\beta_g\,|\,\prod_{i=1}^g\alpha_i \beta_i \alpha_i^{-1}\beta_i^{-1}=1 \rangle.
\end{equation}
The MCG of $\Sigma_g$ can be generated by Dehn twists along $\alpha_i$ and $\beta_i$ for $i=1,\cdots,g$ and $\gamma_i$ for $i=1,\cdots,g-1$. $\gamma_i$ can be written as,
\begin{equation}
  \gamma_i=\alpha_{i+1}^{-1}\tilde{\alpha}_i.
\end{equation}
where $\tilde \alpha_i$ is defined as in \eqref{eq:tilde}. Since our maps for the Dehn twists on the double torus modify only a specific corner
of the polygon while leaving the other parts of it fixed, they generalize naturally to maps on the $4g$-gons. Also as in the previous $g = 2$ case,
to construct $D_{\gamma_i}$, it is easier to work with an irregular $g$-gon. As an example, the $g=3$ case is analyzed in more detail in the Appendix \ref{sec:g3}.

Furthermore, in Appendix \ref{sec:areacalc} we show that this map does not change the area of any region by more than a constant bounded factor, even
in the limit $g \rightarrow \infty$.  As explained in the next section, this feature is important to ensure that the depth of the local geometry change circuit remains constant as one
increases the code distance (see Sec~\ref{sec:reTriangulation}).

\section{Change of triangulation}\label{sec:reTriangulation}

\begin{figure*}[hbt]
\centering
 \includegraphics[width=2\columnwidth]{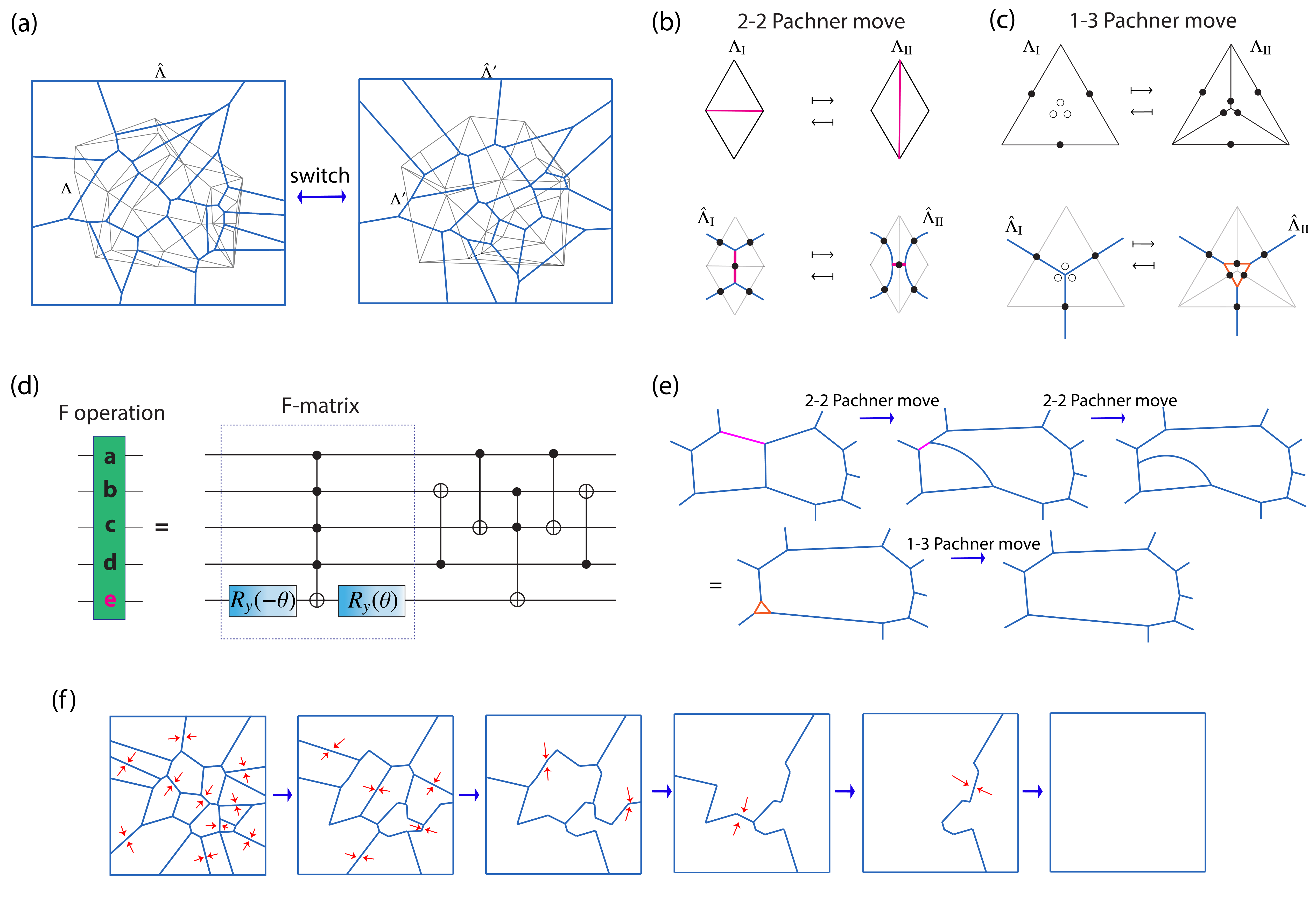}
 \caption{(a) Example of two different triangulations $\Lambda$ and
   $\Lambda'$ (and their dual graphs) with the same average vertex density, which we wish to switch between.
(b,c) Elementary gadgets: 2-2 and 1-3 Pachner moves on the triangulation (represented by $\hat{\Lambda}_I$ to
$\hat{\Lambda}_\text{II}$ in the top row) and its dual trivalent graph (represented by $\hat{\Lambda}_I$ to $\hat{\Lambda}_\text{II}$ in the
bottom row). The black dots represent data qubits, while the white dots represent ancilla qubits initialized to $\ket{0}$.
(d) The quantum circuit implementing the 2-2 Pachner move (F operation) in the Fibonacci model.
(e) The ``edge-sweep" algorithm to merge two edge-sharing plaquettes on a trivalent graph by $s-3$ steps of 2-2 Pachner moves followed
by a 1-3 Pachner move in the end. (f) A parallel merging algorithm. }
  \label{fig:switching}
\end{figure*}

As described in Sec. \ref{sec:geogateset}, step (1) of our protocol permutes the qubits by applying the continuous shear map of
Sec. \ref{sec:continuousMaps} to the triangulation $\Lambda$. After the permutation, the original triangulation $\Lambda$ is
changed to a sheared triangulation $\Lambda'$.  In order to return to the original Hilbert space $\mathcal{H}_{\Lambda}$ and
hence reach a non-trivial unitary map preserving the code space, we need to switch the triangulation back from $\Lambda'$ to $\Lambda$.

In this section, we devise a local unitary circuit to switch a Turaev-Viro code between two arbitrary triangulations
$\Lambda$ and $\Lambda'$. We consider $\Lambda$ and $\Lambda'$ to have the same number of vertices and edges for any
given region, up to at most a constant factor, $c$ [as illustrated in Fig.~\ref{fig:switching}(a)].
Since the switching circuit can be parallelized by acting throughout the whole space at once, the depth of the
circuit only depends on $c$.
To present our algorithm it is more convenient to show the switch between the two dual trivalent graphs instead, as indicated
by the thick blue lines in Fig.~\ref{fig:switching}(a).

For clarity, we drop the branching structure (previously indicated by arrows on the edges) of the graphs in this section.
Many theories of interest, such as the Ising code and the Fibonacci code, do not require the branching structure.

\begin{figure*}[hbt]
\centering
 \includegraphics[width=2\columnwidth]{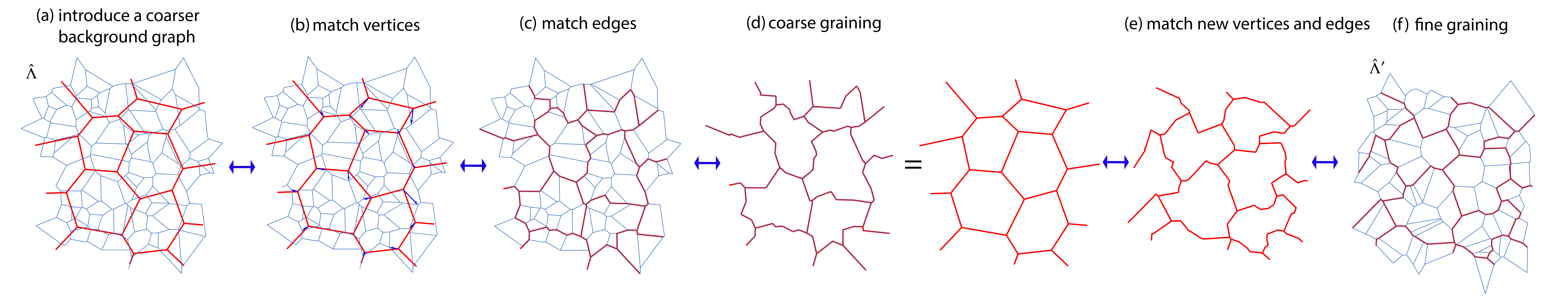}
  \caption{The main switching algorithm between trivalent graphs $\hat{\Lambda}$ and $\hat{\Lambda}'$.}
  \label{fig:main_switching_algorithm}
\end{figure*}

\subsection{Gadgets}

The elementary re-triangulation gadgets we use are associated with the 2-2 and 1-3 Pachner moves, shown in Fig.~\ref{fig:switching}(b) and (c).
They correspond to unitary transformations that take the wave function defined on a triangulation $\Lambda_\text{I}$ to the wave function defined
on a different triangulation $\Lambda_\text{II}$, which differs locally:
\begin{align}
\label{FmoveRel1}
\Psi_{\hat{\Lambda}_\text{I}
}\left(~
\begin{tikzpicture}[baseline={([yshift=-.5ex]current  bounding  box.center)}]
\draw[dualblue,thick,] (0,0) -- (0.4,0);
\draw[dualblue,thick,] (-0.15,0.25) -- (0,0);
\draw[dualblue,thick,] (-0.15,-0.25) -- (0,0);
\draw[dualblue,thick,] (0.55,0.25) -- (0.4,0);
\draw[dualblue,thick,] (0.4,0) -- (0.55,-0.25);
\draw (0.2,0.105) node {$e$};
\draw (-0.175,0.375) node {$b$};
\draw (-0.175,-0.375) node {$a$};
\draw (0.575,0.375) node {$c$};
\draw (0.575,-0.375) node {$d$};
\end{tikzpicture}
\right)
 = \sum_f F^{abc}_{def} \
\Psi_{\hat{\Lambda}_\text{II}}\left(~
\begin{tikzpicture}[baseline={([yshift=-.5ex]current  bounding  box.center)}]
\draw[dualblue,thick] (0,0) -- (0,0.8/2);
\draw[dualblue,thick,] (0,0) -- (0.5/2,-0.3/2) ;
\draw[dualblue,thick,] (-0.5/2,-0.3/2) -- (0,0);
\draw[dualblue,thick,] (0.5/2, 1.1/2) -- (0,0.8/2) ;
\draw[dualblue,thick,] (-0.5/2,1.1/2) -- (0,0.8/2);
\draw (0.24/2,0.4/2) node {$f$};
\draw (-0.8/2,1.25/2) node {$b$};
\draw (-0.8/2,-0.45/2) node {$a$};
\draw ( 0.8/2,1.25/2) node {$c$};
\draw (0.8/2,-0.45/2) node {$d$};
\end{tikzpicture}
\right)
\end{align}
\begin{align}
\label{FmoveRel2}
\Psi_{\hat{\Lambda}_\text{I}}\left(
\begin{tikzpicture}[scale=0.7, baseline={([yshift=-.5ex]current  bounding  box.center)}]
\draw[dualblue,thick,] (0,-2/3.4) -- (0,-2*0.4/3.4);
\draw[dualblue,thick,] (-1/2, 1/3.4 ) -- (-0.2,0.4/3.4);
\draw[dualblue,thick] (1/2,1/3.4) -- (0.4/2,0.4/3.4);
\draw[dualblue,thick,] (0.4/2,0.4/3.4) -- (-0.4/2,0.4/3.4);
\draw[dualblue,thick,] (-0.4/2,0.4/3.4) -- (0,-0.8/3.4);
\draw[dualblue,thick] (0.4/2,0.4/3.4) -- (0,-0.8/3.4) ;
\draw (-1.35/2,0.95/2) node {$b$};
\draw (0,-1.6/2) node {$a$};
\draw (1.35/2,0.95/2) node {$c$};
\draw (0, 0.75/2) node {$d$};
\draw (-0.56/2,-0.12/2) node {$e$};
\draw (0.56/2,-0.12/2) node {$f$};
\end{tikzpicture}
\right)
= [F^{abd}_{fce}]^* \sqrt{\frac{d_d d_f}{d_c}} \
\Psi_{\hat{\Lambda}_\text{II}}\left(
\begin{tikzpicture}[scale=0.4, baseline={([yshift=-.5ex]current  bounding  box.center)}]
\draw[dualblue,thick,] (0,-4/3.4) -- (0,0);
\draw[dualblue,thick,] (-1, 2/3.4 ) -- (0,0);
\draw[dualblue,thick] (1,2/3.4) -- (0,0);
\draw (-1.25,0.8) node {$b$};
\draw (0,-1.7) node {$a$};
\draw (1.25,0.8) node {$c$};
\end{tikzpicture} \right),
\end{align}
where $F^{abc}_{def}$ is the F-symbol that defines the Turaev-Viro code, and $d_c, d_d$, and $d_f$ are the
quantum dimensions of anyons labeled by $c$, $d$ and $f$.  For the 2-2 Pachner move [Eq.~\eqref{FmoveRel1}],
the state labels on the four external legs ($a$,$b$,$c$ and $d$) are fixed, while the internal edge is flipped
(both in the triangulation and trivalent graph) with the state labels changing from $e$ to $f$. The 2-2 Pachner move
can be implemented by unitary gates acting on physical qubits as will be illustrated below.  For the 1-3 Pachner
move [Eq.~\eqref{FmoveRel2}], the state labels on the three external legs ($a$,$b$,$c$ and $d$) are fixed, while a
triangle with three new edges ($e$, $d$ and $f$) is added at the center of the trivalent graph (from
$\hat{\Lambda}_I$ to $\hat{\Lambda}_\text{II}$).  Correspondingly, in the original triangulation, a three-legged
vertex is added in the center of a triangle (from $\Lambda_I$ to $\Lambda_\text{II}$)  [see Fig.~\ref{fig:switching}(c)].
These new edges come from ancilla qubits initialized at $\ket{0}$, which then get entangled into the code by the
1-3 Pachner move, which can be considered as a fine graining procedure. The reverse of this process is a
coarse graining procedure. The 1-3 Pachner move can also be implemented by unitary gates, which can be
decomposed into 2-2 Pachner moves and two other simple unitary gates (see Ref.~\cite{zhu2018, Zhu:2018CodeLong} for details).

In the case of the Fibonacci Turaev-Viro code, we have two simple objects in the unitary fusion category, labeled
$0$ and $1$, with fusion rules $1 \times 1 = 0 + 1$, and the only non-trivial $F$-matrix is:
\begin{align}\label{F-matrix}
F^{111}_{1}=\begin{pmatrix}
\phi^{-1} & \phi^{-\frac{1}{2}}  \\
\phi^{-\frac{1}{2}}&  -\phi^{-1}
\end{pmatrix},
\end{align}
where $\phi=\frac{\sqrt{5}+1}{2}$ is the golden ratio.   All other $F$-symbols are either $1$ or $0$, depending on whether they are consistent
with the fusion rules and Eq.~\eqref{FmoveRel1}. A specific quantum circuit implementing the 2-2 Pachner move ($F$-operations) in the
Fibonacci code was presented in Ref.~\cite{Bonesteel:2012fl} and is shown in Fig.~\ref{fig:switching}(d). The circuit inside the dashed box is
composed of a 5-qubit Toffoli gate sandwiched by two single-qubit rotations, which implements the $F$-matrix in \eqref{F-matrix}.
Here, $R_y(\pm \theta)=e^{\pm i \theta \sigma_y/2}$ represents single-qubit rotations about the y-axis with angle $\theta$$=$$\tan^{-1} (\phi^{-\frac{1}{2}})$.
All the other maps are taken care of by the rest of the quantum circuit in panel (d). For the other widely considered case, the $\mathbb{Z}_2$
hyperbolic  surface codes,  the F-symbols and Pachner moves can be implemented via only CNOTs (see Ref.~\cite{Zhu:2018CodeLong} for details).

Based on these gadgets, we introduce the following two lemmas about the trivalent graphs, which serve as additional gadgets for the main algorithm.
\begin{lemma}\label{lemma1}
  Two edge-sharing plaquettes on a trivalent graph can be merged into a single plaquette using $s-2$ steps of Pachner moves, where
  $s$ is the number of edges in the smaller plaquette (i.e., plaquette with fewer edges).
\end{lemma}
An ``edge-sweeping" algorithm implements the above statement as shown in Fig.~\ref{fig:switching}(e) with $s-3$ steps of 2-2 Pachner moves
and a $1-3$ Pahcner move in the end.

Using the above plaquette-merging gadget, we can also demonstrate the following lemma when considering merging
many plaquettes in parallel, as illustrated in Fig.~\ref{fig:switching}(f). In particular, one merges all mergeable neighboring
pairs in each step.
\begin{lemma}\label{lemma2}
  A collection of $m$ contiguous plaquettes can be merged with $\mathcal{O}\left(\log_2(m)\right)$ rounds of merging
  of the edge-sharing plaquettes. The depth of the algorithm is upper bounded by $\mathcal{O}\left(l \ \log_2(m)\right)$
  steps of Pachner moves. Here, $l$ is chosen to be the number of edges of the largest plaquette in this collection.
\end{lemma}

\subsection{The main algorithm}

Here, we present the main algorithm to switch between the triangulations $\Lambda$ and $\Lambda'$ (and equivalently the dual graphs
$\hat{\Lambda}$ and $\hat{\Lambda}'$).  For the clarity of presentation, we choose a particular order by switching from
$\hat{\Lambda}$ to  $\hat{\Lambda'}$, although the algorithm and corresponding quantum circuit are reversible.
The detailed algorithm is as follows:
\begin{enumerate}
\item
We first introduce a trivalent graph $\hat{\Lambda}_C$ coarser than the two fine trivalent graphs  $\hat{\Lambda}$ (starting graph) and
$\hat{\Lambda}'$ (target graph) we want to switch between. In particular, we consider a coarser graph $\hat{\Lambda}_C$ that encircles at most $m$  plaquettes of the starting finer graph $\hat{\Lambda}$.
We note that $m$ needs to be bounded and independent of code distance $d$. We also require that
any plaquette in the coarser graph $\hat{\Lambda}_C$ contains at least one plaquette of the finer graphs, as illustrated in Fig.~\ref{fig:main_switching_algorithm}(a).

Now the coarser graph $\hat{\Lambda}_C$ also encircles at most
$m'$  plaquettes of the target finer graph $\hat{\Lambda}'$. As discussed in Sec.~\ref{sec:continuousMaps} and Appendix ~\ref{sec:areacalc},
the area ratio for a given infinitesimal code patch before and after the application of the shear maps (belonging to $\hat{\Lambda}$ and $\hat{\Lambda}'$ respectively)
is a bounded constant independent of $d$ (and thus also independent of the number of physical qubits $n$).  This ensures that the ratio $m/m'$ is also bounded and
independent of $d$. Since $m$ is bounded and independent of $d$, $m'$ is also a bounded constant independent of $d$.   This is a
crucial property to ensure the $\mathcal{O}(1)$ depth of the switching circuit and the logical gates.

\item
  We now match the vertices ($v_C$) of the coarser graph $\hat{\Lambda}_C$ with vertices ($v$) on the starting graph $\hat{\Lambda}$ by pinning the vertices of $\hat{\Lambda}_C$ to the closest vertices to them on $\hat{\Lambda}$, as shown in Fig.~\ref{fig:main_switching_algorithm}(b).
  We can now bend each edge ($e_C$) on the coarser graph $\hat{\Lambda}_C$ to match with multiple edges ($e$) on the starting graph, as
  shown in Fig.~\ref{fig:main_switching_algorithm}(c). In this way, each plaquette in the fine graph $\hat{\Lambda}$ is strictly enclosed
  in only one plaquette in the coarser graph $\hat{\Lambda}_C$. This step does not require any quantum
  operations and is done entirely in the classical software.

  Note that the deformation of the coarser graph $\hat{\Lambda}_C$ leads to a slight change of the maximal number of enclosed fine
  plaquettes in any coarser plaquette, i.e., $m$ and $m'$, into $\mathcal{O}(m)$ and $\mathcal{O}(m')$.

\item
  We coarse grain the finer graph $\hat{\Lambda}$ into the coarser graph $\hat{\Lambda}_C$ by using the parallel plaquette
  merging algorithms introduced in  \textbf{Lemma \ref{lemma2}}, as shown in Fig.~\ref{fig:main_switching_algorithm}(d).
  This procedure takes $\mathcal{O}(\log_2(m))$ merging steps with $\mathcal{O}(l \ \log_2(m))$ time steps of Pachner moves, where $l$ is the largest number of edges among all plaquettes in the
  starting finer graph $\hat{\Lambda}$.

\item
We match  vertices ($v_C$) and the edges ($e_C$) of the coarser graph $\hat{\Lambda}_C$ with those ($v'$ and $e'$) in the finer graph $\hat{\Lambda}'$  using the same procedure as above, as shown in Fig.~\ref{fig:main_switching_algorithm}(e).

\item
  We fine grain the coarser graph $\hat{\Lambda}_C$ into the target finer graph $\hat{\Lambda}'$ by reversing the parallel merging
  algorithms in \textbf{Lemma \ref{lemma2}}, as shown in Fig.~\ref{fig:main_switching_algorithm}(f). This procedures takes
  $\mathcal{O}(\log_2(m'))$ splitting steps with $\mathcal{O}(l' \ \log_2(m'))$ time steps of Pachner moves, where $l'$ is the
  largest number of edges among all plaquettes in the target finer graph $\hat{\Lambda'}$.
\end{enumerate}

As we see the total time complexity of the switching algorithm is $\mathcal{O}[\text{max}(l  \log_2(m),l'  \log_2(m'))]$, i.e.,
dominated by the larger complexity from the coarse graining and fine graining process.  Since $m,m',l,l'$ can all be
bounded values independent of the code distance $d$ (or equivalently graph size), we reach the following theorem:

\begin{theorem}
  If two triangulations $\Lambda$ and $\Lambda'$ have bounded ratios in terms of their vertices, edges, and plaquettes per unit area,
  then there exists a bounded depth circuit to convert between them with Pachner moves. The depth of the circuit is independent of the
  area of the surface, and therefore independent of the code distance $d$.
\end{theorem}

We emphasize again that the requirement of the bounded ratios of vertices, edges and plaquettes per unit area is ensured
by the bounded ratio of the areas corresponding to the infinitesimal code patches before and after the shearing maps,
as discussed in Sec.~\ref{sec:continuousMaps} and Appendix \ref{sec:areacalc}.   We also note that in the above discussions,
we focused on proving the existence of a constant depth local unitary circuit, rather than giving the most efficient switching
algorithm.   To be more efficient, one does not need to follow the fine$\rightarrow$coarse$\rightarrow$fine pattern, but
can rather directly find the shortest circuit to directly switch between the two fine graphs via Pachner moves.

\section{Fault tolerance and space-time overhead}\label{sec:FaultTolerance}
So far we have shown that a generating set of Dehn twists for the MCG of a genus $g$ hyperbolic surface can be implemented
by a constant depth unitary circuit, where the depth is independent of the code distance and therefore also the number of physical
and logical qubits, $n$ and $k$. In this section, we briefly discuss fault tolerance of these circuits.

Our circuit breaks up into two basic pieces: A permutation of the physical qubits and a local constant depth circuit that implements the retriangulation. Since local unitary circuits have a linear light cone, the latter, i.e. the local constant depth circuit, is intrinsically fault tolerant. So, to ensure the fault tolerance of our procedure, we concentrate on the first part which permutes the physical qubits.

The permutation in our maps requires qubits to be permuted over long distances. Due to its non-local nature, there are two main concerns in regard to the propagation of errors that we need to address: (1) What happens to the pre-existing local error strings? Is it possible for them to be enlarged to lengths of $\mathcal{O}(d)$? (2) Is it possible to introduce new non-local ($\mathcal{O}(d)$ long) error strings by a noisy implementation of the permutation circuit, e.g. noisy SWAP gates?

For a generic non-local permutation, both issues mentioned above could possibly arise. Nevertheless the permutations that we
utilize have a special structure. To address the first issue, note that the continuous maps introduced in Sec. \ref{sec:continuousMaps}, take
two points which are $\mathcal{O}(1)$ apart to new points which are still $\mathcal{O}(1)$ apart. This can be seen from the analysis carried out in Appendix \ref{sec:areacalc} as well. However, to enlarge a local error string to a large error string of length $\mathcal{O}(d)$ we have to separate its endpoints by a factor of $\mathcal{O}(d)$. Thus, we can conclude that after the permutation, all pre-existing local error strings would remain local; at worst their length will be increased by a constant factor independent of $d$.
Stated differently, our constant depth circuits map any operator with support in a spatial region $\mathcal{R}$ to
an operator with support in a spatial region $\mathcal{R'}$, where the area of $\mathcal{R}$ and $\mathcal{R'}$ are related by a constant (independent
of code distance) factor. Here the areas are with respect to the hyperbolic metric. In this sense, the whole circuit is also
``locality-preserving,'' although strictly speaking the phrase ``locality-preserving'' is often reserved for the case where the constant
factor is unity\cite{Bravyi:2013dx, Beverland:2016bi}..

Now we consider the possibility of introducing new non-local error strings during a noisy implementation of the permutation. Let's say the permutation is implemented by a set of noisy long range SWAPs. The important point to note is that while the SWAP operations are long ranged, they are still \textit{low weight} operators. In particular, each SWAP operation acts on $2$ qubits. On the other hand, a logical error string is a high weight operator, consisting of $\mathcal{O}(d)$ single qubit errors. So, if we assume the errors occur independently on different SWAPs, the possibility of introducing a logical error by a set of noisy SWAPs is still exponentially small in $d$. Therefore the second question can be answered in the negative as well.

Therefore, our Dehn twists and the corresponding logical gates are inherently protected from errors, in the sense that they do not stretch
error strings by more than a constant factor, nor can they introduce error strings that have length more than a constant, independent
of code distance.

However, if we apply the same collection of Dehn twists (logical gates) repetitively in the same region of the manifold,
in the worst case the length of error strings could grow exponentially with the number of logical gates being applied. In the absence of measurement noise, the error string can be decoded and corrected in $\mathcal{O}(1)$ time \cite{Burton:2017gr} immediately after the application of a single logical gate (here we have ignored the classical computation time of the decoder which still typically scales with system size). Hence the computation scheme will have an $\mathcal{O}(1)$ (constant) time overhead.
However, in the presence of measurement noise, the error string cannot be immediately decoded and corrected in $\mathcal{O}(1)$
time, so the growth of such a string would be inevitable.   After performing $\mathcal{O}(\log d)$ logical gates in the
same region without any measurement or error correction in between, the error string may grow to a length of $\mathcal{O}(d)$,
which will cause the decoder to fail. Therefore, one has to insert $\mathcal{O}(d)$ rounds of measurements, decoding,
and error corrections for every $\mathcal{O}(\log d)$ of logical gates in the same region. This suggests a  sub-linear overhead
$\mathcal{O}(d/\log d)$ in the computational time when repetitively applying logical gates in the same region, if the measurement
error is taken into account. It may be possible to further reduce such overhead by some additional tricks, at least for
certain types of logical circuits, but there may still be such a sub-linear overhead in the most generic situation.
We note that the above $\mathcal{O}(d/\log d)$ time overhead is an estimate suggested by the considerations stated above;
further work is required to develop efficient decoders to concretely demonstrate the validity of this estimate.

The above statements for the asymptotic space-time resource costs have assumed that the codes have a finite error threshold
and that an efficient decoder exists and requires $\mathcal{O}(d)$ rounds of syndrome measurements to decode errors.
For the Ising anyon code, this has been confirmed in the recent theoretical study in the presence of measurement
error \cite{Dauphinais:2017bz}. For the Fibonacci anyon code, a decoding scheme showing a finite error threshold has only
been developed for the case without measurement noise \cite{Burton:2017gr}. We expect that the Fibonacci anyon code can also be used for fault-tolerant quantum
computation using $\mathcal{O}(d)$ rounds of error correction in the presence of measurement errors. However the
development of concrete decoding algorithms, explicit demonstrations of a finite error threshold, and efficient error
recovery operations for the Fibonacci anyon code requires further work.

\section{Discussion}
\label{sec:conclusion}

We have shown that Dehn twists in hyperbolic Turaev-Viro codes, which are the most general types of topological codes,
can be implemented by constant depth unitary circuits while preserving the constant space overhead of the codes.
We have presented explicit representative maps for Dehn twists on hyperbolic surfaces, which we use to define the
first step of our protocols, which correspond to the permutation of the qubits. As we explicitly verify, our
maps do not increase or shrink any infinitesimal area by more than a constant factor even in the $g \rightarrow \infty$
limit. This fact implies that the second step of our protocol, the local unitary circuit that implements the retriangulation back
to the original code space, is of constant depth. Since generic permutations can be implemented as a depth-2 circuit consisting
of long-range SWAP operations applied in parallel, it follows that our entire protocols are constant depth unitary circuits.

Because our circuits for a given Dehn twist only act within a local region associated with a given handle or pair of handles,
they can be applied in different regions simultaneously, in parallel (again while still keeping the constant space overhead), which
makes our protocols amenable to parallel quantum computation.

The simplest example of the codes that we study is the $\mathbb{Z}_2$ hyperbolic surface code. Our results demonstrate
that Dehn twists, which generate a subgroup of the Clifford group acting on the logical qubits in this case, can be implemented
by constant depth unitary circuits. In Appendix \ref{sec:surface} we demonstrate explicitly how our retriangulation circuit in this case
can be implemented using CNOT operations on the physical qubits.

When applying our results to the Fibonacci Turaev-Viro code, our results imply that a universal set of gates can be implemented through
constant depth unitary circuits in a code with finite encoding rate (constant space overhead).  Recent numerical study has
shown the existence of a finite error threshold for a decoding scheme in the Fibonacci anyon code without considering measurement noise \cite{Burton:2017gr}.
We expect, though it has not yet been explicitly demonstrated, that the Fibonacci Turaev-Viro code possesses a finite error threshold,
efficient decoding algorithms and error recovery procedures even in the presence of measurement noise (similar statements
for other non-Abelian codes have been demonstrated \cite{Dauphinais:2017bz}).
Assuming this, and inserting $\mathcal{O}(d)$ rounds of error correction
for every $\mathcal{O}(\log d)$ logical gates to ensure fault-tolerance in the presence of noisy syndrome measurements, our results
suggest that universal fault-tolerant quantum computation with constant space overhead and time overhead of $\mathcal{O}(d/\log d)$
per logical gate is possible. This suggests that a generic logical quantum circuit of depth $D$ with $k$ logical qubits would
require $\mathcal{O}(d D/\log d)$ time steps and $\mathcal{O}(k)$ physical qubits.

To our knowledge, the only proposed scheme with comparable space-time overhead for universal quantum computation
is the one proposed by Gottesman in Ref.~\cite{Gottesman:2014ug} which also uses constant space overhead.
In Gottesman's scheme, the time overhead per logical gate is constant for \it sequential \rm quantum circuits, where one logical gate is applied at each time step.
This means that a highly parallel circuit of depth $D$ would take time $\mathcal{O}(kD)$ \footnote{Ref. \cite{Gottesman:2014ug} also
  proposed methods to reduce this time cost to $\mathcal{O}(k^\gamma D)$ for $\gamma > 0$ arbitrarily small, at the cost of increasing
  the scaling of the space overhead.}.

In the regime of interest for large-scale universal fault-tolerant quantum computation, $k \gg d$, and therefore our
scheme would have an order $\mathcal{O}(d/(k\log d))$ improvement over the proposal of Ref.~\cite{Gottesman:2014ug} for parallel quantum circuits.
We note that $k \gg d$ is also the natural setting for the hyperbolic code,  since $d \sim \log n \sim \log k$ suggests
an exponential growth of logical qubit number $k$ as a function of the code distance $d$.

Our protocols allowing for implementation of mapping class group elements via constant depth circuits
should be straightforwardly generalizable to higher dimensional topological codes. We note that a number
of codes in three and four dimensions (such as the four-dimensional toric code) allow for single-shot error
correction \cite{Bombin:2015hia}. This would potentially allow any number of our type of logical gates
to be applied with constant time overhead, even in the presence of measurement errors.

In this paper we studied hyperbolic codes for two-dimensional topological states, where $d \sim \log n$.
However better codes exist, such as the hypergraph product codes \cite{Tilich2014, Leverrier:2015ju}
or homological codes associated with four-dimensional hyperbolic manifolds \cite{Guth:2014cj}. These
codes also have constant space overhead (and finite error thresholds because they are LDPC), but with the
improved scaling of $d \sim \sqrt{n}$ for the hypergraph product codes and $d \sim n^\epsilon$ ($\epsilon<0.3$)
for the four-dimensional hyperbolic codes. However, due to Mostow rigidity, the mapping class
group of higher dimensional hyperbolic manifolds is necessarily finite. It is therefore a fundamental
question whether there exist codes with constant space overhead and $d \sim n^\alpha$ (with $\alpha > 0$),
and which also admit a universal logical gate set through constant depth unitary circuits.

\section{Acknowledgments}

We thank M. Freedman, J. Haah, and M. Hastings for discussions. We are especially grateful to M. Freedman for helpful suggestions
regarding the retriangulation protocols. This work is supported by NSF CAREER (DMR-1753240), Alfred P. Sloan Research Fellowship, and
JQI- PFC-UMD. G.Z. was also supported by ARO-MURI.

\appendix
\section{Dehn Twist}\label{sec:dehn}
\begin{figure}[t]
  \includegraphics[width=\columnwidth]{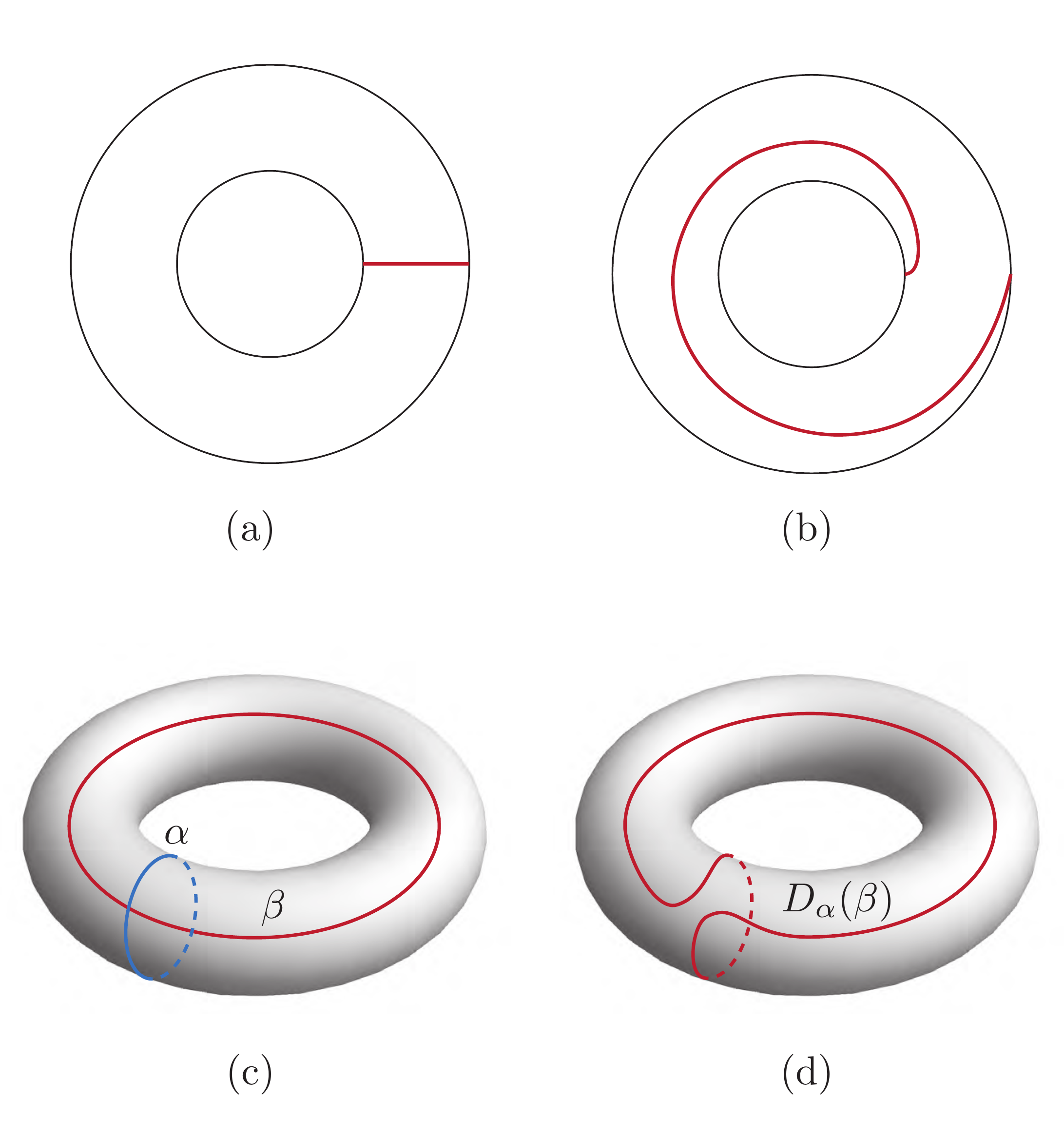}
  \caption{(a) Annulus $A$ with a typical line connecting its boundaries together. (b) Image of the region $A$ under the twist map. (c) and (d) show the $\beta$ loop and its image under the Dehn twist around the $\alpha$ loop, respectively. }
  \label{fig:dehn}
\end{figure}
Consider the annulus $A$ shown in Fig. \ref{fig:dehn}a which consists of the points in the $(r,\theta)$-plane with $1\le r \le 2$ and $0\le \theta < 2\pi$. Let $T:A \to A$ denote the twist map given as:
\begin{equation}
  T(r,\theta)=\qty(r,\theta+ 2\pi(r-1)).
\end{equation}
Note that $T$ is an orientation preserving homeomorphism which reduces to the identity map on the boundaries of $A$. Fig. \ref{fig:dehn}b shows the image of a line in $A$ under $T$.

Now, let $\Sigma$ denote an arbitrary oriented surface and let $\alpha$ be a simple closed curve on $\Sigma$. To define the Dehn twist around $\alpha$, first we choose a regular neighborhood $\mathcal{A}$ of $\alpha$ on $\Sigma$ which is homeomorphic to $A$. Let $\phi:A\to \mathcal{A}$ denote an orientation preserving homeomorphism from $A$ to $\mathcal{A}$. We define the twist map around $\alpha$ by the following homeomorphism $T_\alpha:\Sigma \to \Sigma$
\begin{equation}\label{eq:}
  T_\alpha(x)=
  \begin{cases}
  \phi~\circ T\circ \phi^{-1}(x)\quad & x \in \mathcal{A}\\
  x & x \notin \mathcal{A}
  \end{cases}
\end{equation}
$T_\alpha$ clearly depends on the choice of $N$, $\phi$ and $\alpha$. We define the \textit{Dehn Twist} around $\alpha$, denoted by $D_\alpha$ to be the isotopy class of $T_\alpha$, i.e. the class of homeomorphisms of $\Sigma$ to itself that can be deformed continuously to $T_\alpha$. We remark that $D_\alpha$ depends only on the isotopy class of $\alpha$.

As an example, one can consider the Dehn twist around the $\alpha$ loop on the simple torus shown in Fig.\ref{fig:dehn}c. Figure \ref{fig:dehn}d shows how the $\beta$ loop gets deformed by the action of $D_\alpha$.

A simple way to find the image of a given loop like $\beta$ under the action of the Dehn twist around another loop like $\alpha$, called the \text{Dehn surgery}, is as follows: We start by tracing out the $\beta$ loop until we hit an intersection with the $\alpha$ loop. Then we turn left to trace out the $\alpha$ loop until we return to the intersection point, where we turn right to continue tracing out the $\beta$ loop. We need to do the same at any intersection of $\alpha$ and $\beta$ until we get back to the starting point. Note that with the above definition the Dehn twist around $\alpha$ does not depend on any direction that the $\alpha$ loop might have.

\section{Area scaling calculation}\label{sec:areacalc}
In this appendix we analyze the shearing maps introduced in Section \ref{sec:continuousMaps} to see by
how much they scale the area locally. Because these maps do not have any singularity on a genus $g$
surface, the local area scaling is bounded from below and above as one considers a fixed $g$. But more
importantly, we will show that if we consider a family of maps on surfaces with increasing genus $g$, the
local area scaling remains finite and bounded from below. Note that this result is crucial for the
retriangulation circuit described in Sec. \ref{sec:reTriangulation} to be constant depth.

As in the rest of the paper, we use the Poincare disk model for $\mathbb{H}^2$, namely the open unit disk in $\mathbb{C}$ with the Reimannian metric:
\begin{equation}\label{eq:metric}
  \dd s^2 =4\frac{\dd x^2 + \dd y^2}{(1-r^2)^2}.
\end{equation}
In this model, the geodesics connecting two points would be either circular arcs perpendicular to the unit disk or straight lines passing through the origin. It would be useful to note that a line reflection in $\mathbb{H}^2$ looks like circle inversion in the unit disk model.

\begin{figure}[!t]
  \includegraphics[width=\columnwidth]{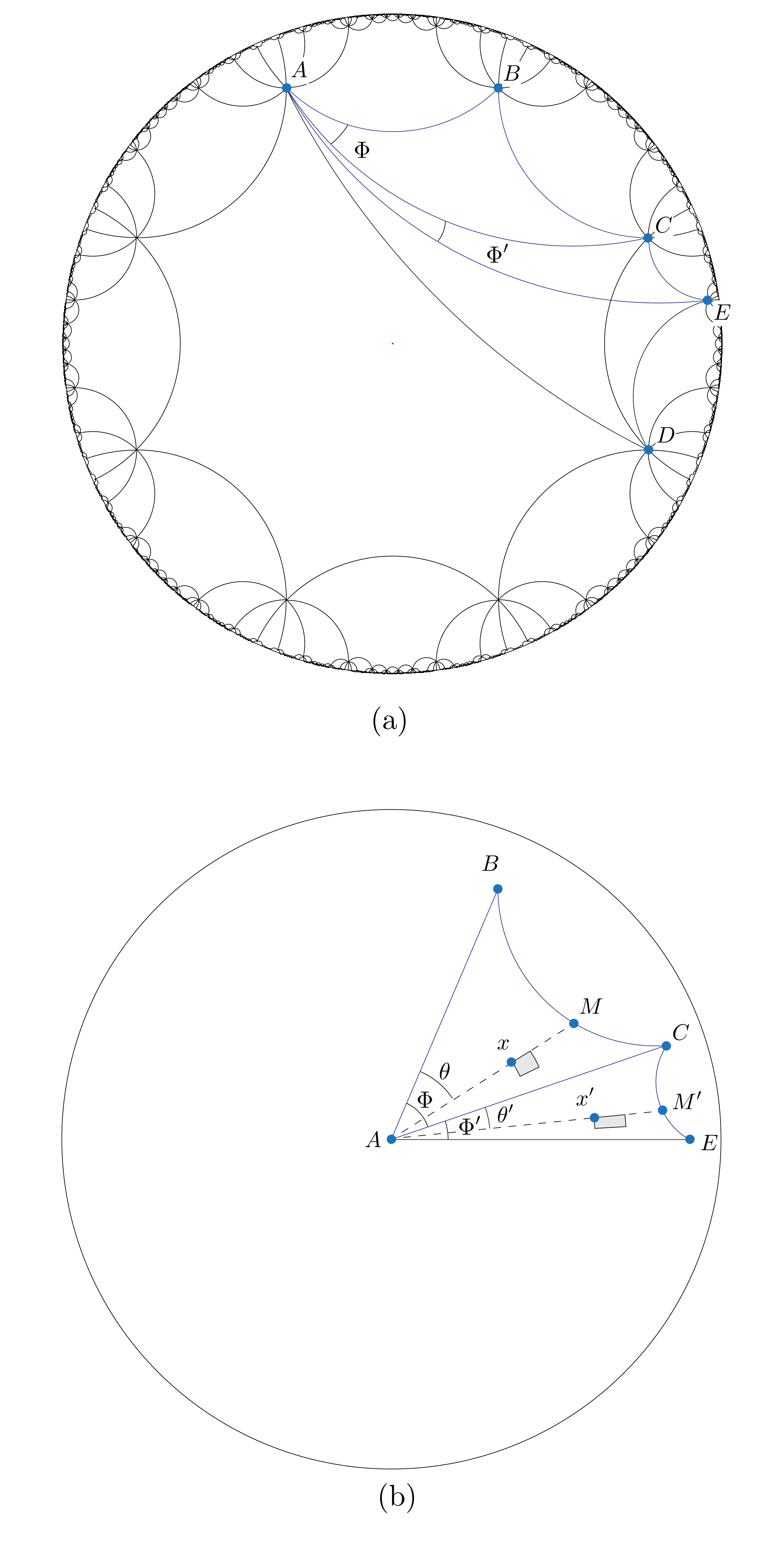}
  \caption{(a) $D^\ast_{\alpha_1}$ maps the $\Delta ABC$ and $\Delta ACD$ triangles to the $\Delta ACE$ and $\Delta AED$ triangles respectively. (b) By moving the origin of the Poincar\`e disk to the point $A$, the geodesics $AB$, $AC$ and $AE$ turn into straight lines while the angles between them remains the same. A typical area element at point $x$ alongside its image at $x'$ is shown on the figure.}
  \label{fig:areacalc1}
\end{figure}

A genus $g$ hyperbolic surface can be obtained by compactifying a canonical $N$-gon for $N=4g$. A canonical $N$-gon plotted in the
Poincare disk model and centered at the origin would have its vertices at points:
\begin{equation}
  z_j=\sqrt{\cos(\frac{2\pi}{N})} e^{i 2\pi j/N},\quad j=0,\cdots,N-1.
\end{equation}
The length of the sides, as measured using the hyperbolic metric, would be (see Fig.\ref{fig:areacalc1}a):
\begin{equation}\label{eq:AB_exact}
  |AB|=4\tanh(\frac{\cos(\pi/N)-\sin(\pi/N)}{\sqrt{\cos(2\pi/N)}}).
\end{equation}
 The $\{N,N\}$-tiling of $\mathbb{H}^2$ can be obtained by reflecting the canonical $N$-gon with respect to its sides and repeating this procedure indefinitely.

 We start by analyzing the $\alpha_1$ Dehn twist in detail. With minor modifications, the same calculation applies to the other maps as well,
 so we will only mention the end results for the other maps.

For the points in the shaded region but outside the $ABCD$ hyperbolic $4$-gon in Fig. \ref{fig:dalpha}b, $D_{\alpha_1}$ acts as
identity and hence it is preserves area locally. Next we consider the points inside $\Delta ABC$ which are mapped to
points inside $\Delta ACE$ (See Fig.\ref{fig:areacalc1}).

We start by calculating some lengths and angles. As in the main text, $|~|$ denotes the hyperbolic length.
$|AC|$ can be obtained using the hyperbolic law of cosines (see Fig.\ref{fig:areacalc1}a):
\begin{equation}\label{eq:AC_exact}
  \cosh(|AC|)=\cosh(|AB|)^2-\sinh(|AB|)^2\cos(2\pi/N),
\end{equation}
where we used the fact $|BC|=|AB|$. Then, we can calculate the angle $\Phi\equiv \angle BAC=\angle BCA$ by the hyperbolic law of sines:
\begin{equation}\label{eq:phi_exact}
  \sin(\Phi)=\frac{\sinh(|AB|)}{\sinh(|AC|)}\sin(2\pi/N).
\end{equation}
Note that $|CE|=|AB|$ and $\angle ECD=2\pi/N$. Therefore, $|AE|$ and $\Phi'\equiv \angle CAE $ can be computed using the hyperbolic laws of cosines and sines respectively:
\begin{align}
  &\cosh(|AE|)=
  \begin{aligned}[t]
    &\cosh(|AB|)\cosh(|AC|)\\
                &-\sinh(|AB|)\sinh(|AC|)\cos(4\pi/N-\Phi)
  \end{aligned}\label{eq:AE_exact}\\
 &\sin(\Phi')=\frac{\sinh(|AB|)}{\sinh(|AE|)}\sin(4\pi/N-\Phi).\label{eq:phi_p_exact}
\end{align}
To make calculations simpler, we place the origin of the Poincar\`e disk on point $A$; by doing so, the geodesics that come out of the point $A$ will look like straight lines on the disk. In Fig. \ref{fig:areacalc1}b, $\Delta ABC$ and $\Delta ACE$ are plotted in this new coordinate frame.

Consider an infinitesimal area element $\dd s$ at $x=(r,\theta)$ inside $\Delta ABC$ which is mapped to the area element
$\dd s'$ at $x'=(r',\theta')$ inside $\Delta ACE$; $r$ and $r'$ are the Euclidean distance from origin in Fig. \ref{fig:areacalc1}b
and the $\theta$ and $\theta'$ angles are measured with respect the lines $AB$ and $AC$ respectively. we define $\lambda(r,\theta)$ as the scale factor of the area:
\begin{equation}\label{eq:AreaRatio}
  \lambda_N(r,\theta)\equiv \frac{\dd s'}{\dd s}=\sqrt{\frac{g(r')}{g(r)}}\frac{r'\dd r'}{r\dd r}\frac{\dd \theta'}{\dd \theta},
\end{equation}
where $g(r)$ is the determinant of the hyperbolic metric:
\begin{equation}\label{eq:gr}
  g(r)=\frac{16 r^2}{(1-r^2)^4}.
\end{equation}
We want to show that $\lambda$ is bounded from above and below:
\begin{equation}\label{eq:bcon}
  L<\lambda_N(r,\theta)<U,
\end{equation}
for some real positive $L$ and $U$. First, we need to find $r'$ and $\theta'$ in terms of $r$ and $\theta$.

Using the law of cosines for the angles in $\Delta ABM$ (see Fig. \ref{fig:areacalc1}b) gives:
\begin{align}
  \cos( \angle AMB)=&-\cos(\theta)\cos(\frac{2\pi}{N})\nonumber\\
  &+\sin(\theta)\sin(\frac{2\pi}{N})\cosh(|AB|)
\end{align}
And we can use this result to find $|BM|$ and $|AM|$:
\begin{align}
  &\sinh(|AM|)=\frac{\sin(2\pi/N)}{\sin(\angle AMB)}\sinh(|AB|)\label{eq:AM_exact}\\
  &\sinh(|BM|)=\frac{\sin(\theta)}{\sin(\angle AMB)}\sinh(|AB|)
\end{align}
$D^\ast_{\alpha_1}$ is defined such that:
\begin{align}
  |CM'|=\frac{|BM|}{|BC|}|CE|.
\end{align}
Note that in this case $|BC|=|CE|$ and hence:
\begin{equation}
  |CM'|=|BM|.
\end{equation}
We can find $|AM'|$ by using the law of cosines in $\Delta ACM'$:
\begin{align}
  \cosh(|AM'|)=&\cosh(|CM'|)\cosh(|AC|)\label{eq:AM_p_exact}\\&-\sinh(|CM'|)\sinh(|AC|)\cos(\frac{4\pi}{N}-\Phi).
\end{align}
It is then straight forward to find $r'$ and $\theta'$. According to the way the map $D^\ast_{\alpha_1}$ is defined, we have:
\begin{equation}
  |Ax'|=\frac{|AM'|}{|AM|}|Ax|.
\end{equation}
Note that $r'=\tanh(|Ax'|/2)$ and $|Ax|=2\arctanh(r)$. So we have:
\begin{equation}\label{eq:rpr}
  r'=\tanh(\rho \arctanh(r)),
\end{equation}
where $\rho$ is equal to $\frac{|AM'|}{|AM|}$ and depends only on $\theta$(not $r$). By taking the derivative of \eqref{eq:rpr} with respect to $r$ we get:
\begin{equation}\label{eq:drpdr}
  \frac{\dd r'}{\dd r}=\rho(\theta)\frac{1-r'^2}{1-r^2}
\end{equation}
$\theta'$ can also be obtained by using the hyperbolic law of sines in $\Delta ACM'$:
\begin{equation}\label{eq:theta_p_exact}
  \sin(\theta')=\frac{\sinh(|CM'|)}{\sinh(|AM'|)}\sin(\frac{4\pi}{N}-\Phi)
\end{equation}
By plugging \eqref{eq:rpr},\eqref{eq:drpdr} and \eqref{eq:gr} into \eqref{eq:AreaRatio} we find:
\begin{widetext}
  \begin{equation}\label{eq:lambdaexp}
    \lambda_N(r,\theta)=\qty(\frac{\tanh(\rho \arctanh(r))}{r})^2 \qty(\frac{1+r}{1+\tanh(\rho \arctanh(r))}) \qty(\frac{1-r}{1-\tanh(\rho \arctanh(r))})\rho(\theta)\frac{\dd \theta'}{\dd \theta}.
  \end{equation}
\end{widetext}

First, we fix $\theta$ and see how $\lambda$ changes as one varies $r$ in the range $[0,r_\text{max}=\tanh(|AM|/2)]$. It is straight forward to show that  $\tanh(\rho \arctanh(r))/r$ is bounded by $1$ and $\rho$. The second parentheses in \eqref{eq:lambdaexp}, which is equal to $(1+r)/(1+r')$, also is bounded by $1/2$ and $2$. The third parenthesis in \eqref{eq:lambdaexp} is a monotonic function of $r$, as one can verify by taking the derivative, and so is bounded by $1$ and $(1-\tanh(|AM|/2))/(1-\tanh(|AM|'/2))$. Therefore, to make sure that $\lambda_N(r,\theta)$ is bounded, it suffices to show that $\rho(\theta)=\frac{|AM|'}{|AM|}$
, $(1-\tanh(|AM|/2))/(1-\tanh(|AM|'/2))$ and $\frac{\dd \theta'}{\dd \theta}$ remain finite as one changes $\theta$ from $0$ to $\Phi$.

So far, all expressions were exact. But, since we are interested in the $N\to\infty$ limit, the calculation can be simplified greatly by computing the large $N$ expansion of each expression. In particular, \eqref{eq:AB_exact},\eqref{eq:AC_exact},\eqref{eq:AE_exact},\eqref{eq:phi_exact} and \eqref{eq:phi_p_exact} have the following asymptotic forms:
\begin{align}\label{eq:assympt1}
  &|AB|=-2\ln\qty(\frac{\pi}{2N})-\frac{7\pi^2}{6N^2} -\frac{487 \pi^4 }{720 N^4}+\mathcal{O}(1/N^5)\nonumber\\
  &|AC|=-2\ln\qty(\frac{\pi}{4N})-\frac{61 \pi^2}{24 N^2} +\mathcal{O}(1/N^3)\nonumber\\
  &|AE|=-2\ln\qty(\frac{\pi}{14N})+\mathcal{O}(1/N)\nonumber\\
  &\Phi=\frac{\pi}{2N}+\frac{3\pi^3}{8N^3}+\mathcal{O}(1/N^4)\nonumber\\
  &\Phi'=\frac{\pi}{14 N}+\mathcal{O}(1/N^2)
\end{align}
To find the asymptotic form of functions that involve $\theta$, first we trade $\theta$ for $\eta\equiv \frac{\theta}{\Phi}$. The reason is that $\theta$ varies in the range $[0,\Phi]$ and so has an implicit $1/N$ dependence because of $\Phi$. By using $\eta$ instead of $\theta$, the only small parameter of our expressions would be $1/N$. Note that $0 \le \eta \le 1$. In terms of $\eta$, we get the following asymptotic forms for the expressions in \ref{eq:AM_exact},\ref{eq:AM_p_exact} and \ref{eq:theta_p_exact}:
\begin{align}
  &|AM|=-\ln(\frac{\pi\sqrt{\eta(1-\eta)}}{4 N})+\mathcal{O}(1/N)\\
  &|AM'|=-\ln(\frac{\pi\sqrt{\eta(1-\eta)}}{(16+33\eta) N})+\mathcal{O}(1/N)\\
  &\theta'=\frac{7\pi \eta }{(32+66\eta)N}+\mathcal{O}(1/N^2)
\end{align}
The first two expressions are only valid for $0<\eta<1$; for $\eta=0,1$ we have to use the expressions listed in \eqref{eq:assympt1} instead. Note that for $\eta=0$, we have $|AM|=|AB|$ and $|AM'|=|AC|$. Similarly, $|AM|=|AC|$ and $|AM'|=|AE|$ when $\eta=1$. It follows then,
\begin{align}\label{eq:limits1}
  &\lim_{N\to\infty }\rho(\eta)=1\nonumber \\
  &\lim_{N\to\infty} \frac{1-\tanh(|AM|/2)}{1-\tanh(|AM|'/2)}=\frac{4}{16+33\eta}\nonumber\\
  &\lim_{N\to\infty}\frac{\dd \theta'}{\dd \theta}=\frac{122}{(16+33\eta)^2},
\end{align}
and clearly all of them are bounded as a function of $\eta$. Thus we conclude that the inequality \eqref{eq:bcon} holds and therefore, when mapping $\Delta ABC$ to $\Delta ACE$, local area scaling is finite and bounded form above and belllow.

It remains to show that the same holds when mapping $\Delta ACD$ to $\Delta AED$. The steps are quite the same. First we calculate the side lengths and angles of these triangles. Note that $|DE|=|AC|$,$|CD|=|AB|$ and $\angle ACD=2\pi/N-\Phi$. $|AD|$ and $\Phi''\equiv \angle CAD$ can then be obtained using the hyperbolic law of cosines and sines respectively:
\begin{align}
  &\cosh(|AD|)=
  \begin{aligned}[t]
    &\cosh(|AC|)\cosh(|AB|)\\
    &-\sinh(|AC|)\sinh(|AB|)\cos(\frac{2\pi}{N}-\Phi),
  \end{aligned}\\
  &\sin(\Phi'')=\frac{\sinh(|AB|)}{\sinh(|AD|)}\sin(\frac{2\pi}{N}-\Phi).
\end{align}
Let $M$ be a point on the side $CD$ and $M'$ its image which will be on $ED$. According to the definition of the map,
\begin{equation}
  |EM'|=\frac{|DE|}{|CD|}|CM|=\frac{|AC|}{|AB|}|CM|
\end{equation}
We define $\rho$, $\theta$, $\theta'$, and $\eta$ similar to the previous case: $\theta\equiv \angle MAC$, $\theta'\equiv \angle M'AE$, $\rho(\theta) \equiv |AM'|/|AM|$ and $\eta \equiv  \theta/\Phi''$. As in the previous section, we only need to compute the corresponding limits listed in \eqref{eq:limits1}. The calculation follows the same steps and at the end we will find that:
\begin{align}
  &\lim_{N\to\infty }\rho(\eta)=1 \\
  &\lim_{N\to\infty} \frac{1-\tanh(|AM|/2)}{1-\tanh(|AM|'/2)}=\frac{8}{49-45\eta}\\
  &\lim_{N\to\infty}\frac{\dd \theta'}{\dd \theta}=\frac{112}{(49-45\eta)^2}.
\end{align}
Since all of them are bounded for $0 \le \eta \le 1$ it follows that the area elements in $\Delta ACD$ also expand or shrink by a finite factor under this transformation.

Putting everything together, we conclude that $D^\ast_{\alpha_1}$ has finite local area scaling over the entire polygon and this ends the proof.

The situation for $\beta$ Dehn twists are exactly the same up to some relabeling and needs no more analysis. One can also carry out similar calculations to check that the same remains true for the $\gamma$ Dehn twists. Note that $D^\ast_\gamma$ is essentially the same as $D^\ast_\alpha$; the only difference is that the polygon is no longer regular.

\section{Dehn Twists on $\Sigma_3$}\label{sec:g3}

\begin{figure}[t]
  \includegraphics[width=\columnwidth]{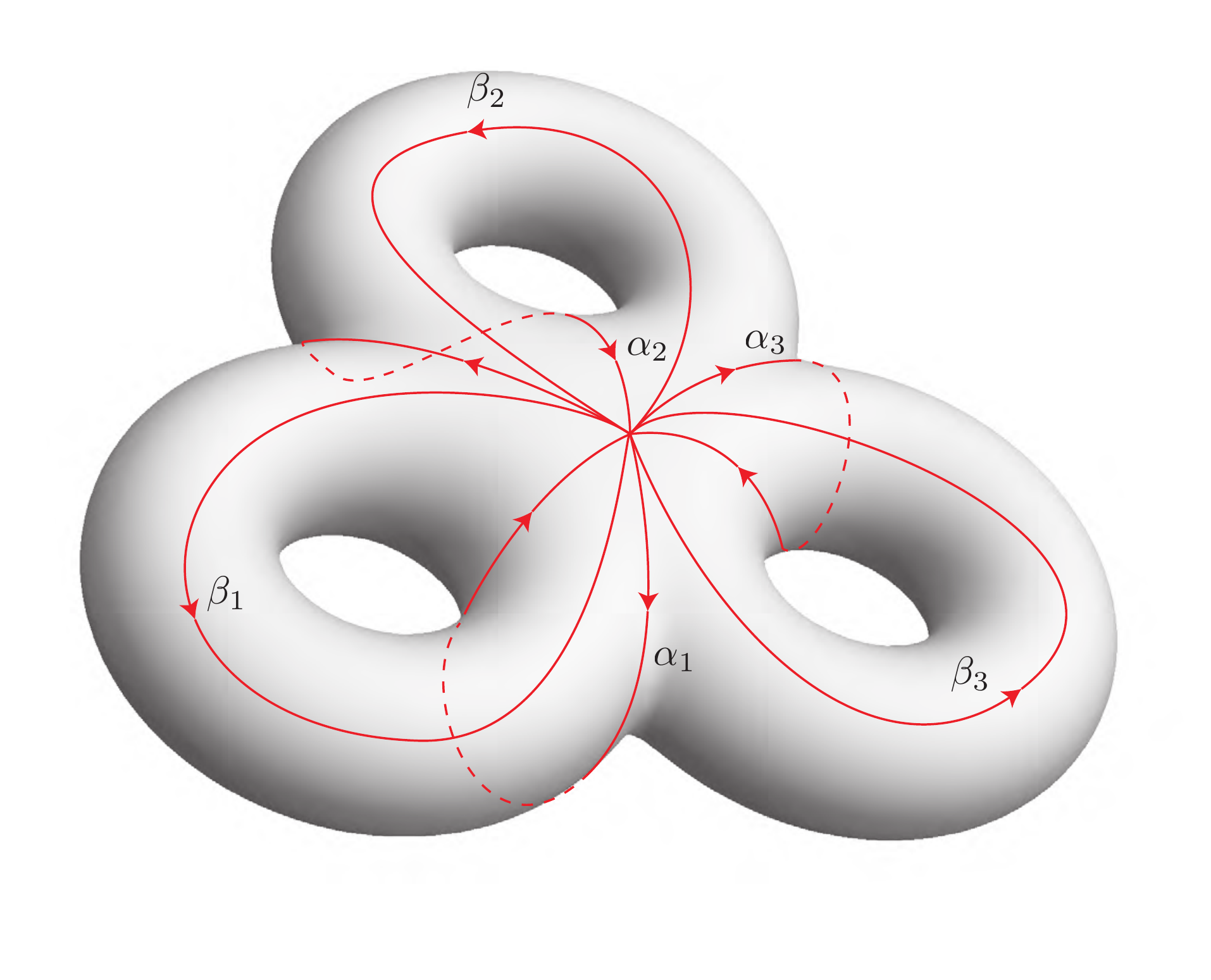}
  \caption{$\Sigma_3$ surface and generators of its fundamental group.}
  \label{fig:sigma3}
\end{figure}
\begin{figure*}[!t]
  \includegraphics[width=0.9\textwidth]{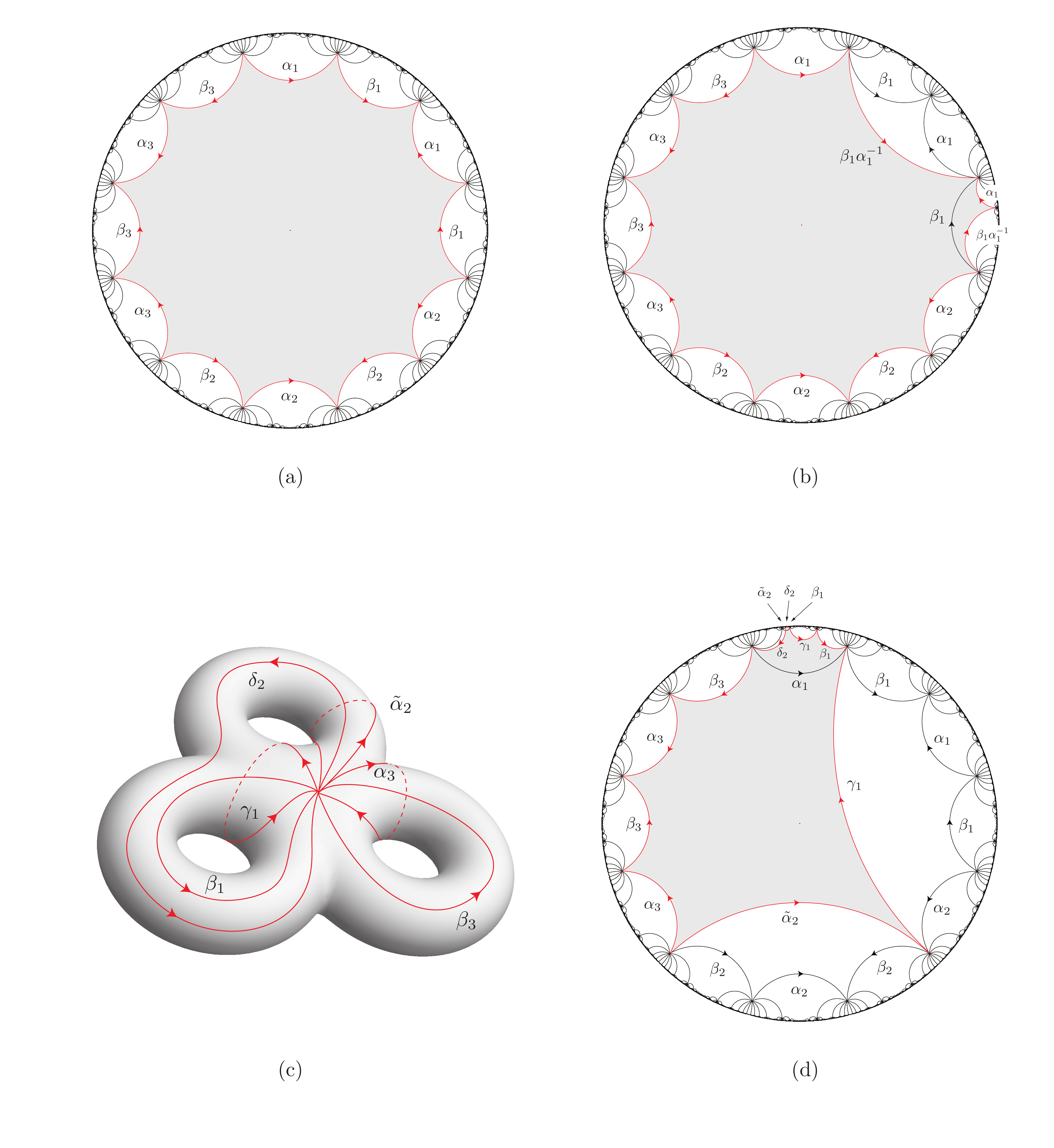}
  \caption{(a) The regular $12$-gon corresponding to the $\Sigma_3$
    surface. (b) The image of the fundamental domain shown in (a) under the action of $D^\ast_{\alpha_1}$. (c) New set of loops on
    the surface of $\Sigma_3$ (d) The fundamental domain that is used to define $D^\ast_{\gamma_1}$ which can be obtained by cutting the
    $\Sigma_3$ surface along the loops shown in (c). The sides associated with $\tilde{\alpha}_2$, $\beta_1$, $\delta_1$ marked
    by the arrows could not be drawn completely due to size constraint.}
  \label{fig:g3new}
\end{figure*}

Although we used a $g = 2$ hyperbolic surface to describe our procedure, as we mentioned in the main text the protocols can
be easily generalized to higher genus surfaces. In this section, we briefly explain how the generalization applies in the
$g = 3$ case. The genus $3$ surface $\Sigma_3$, shown in Fig. \ref{fig:sigma3}, can be constructed by identifying every
other edge of a $12$-gon (see Fig. \ref{fig:g3new}a). The space of possible hyperbolic metrics -- Teichm\"uller space -- corresponds
to inequivalent choices for the locations of the vertices of the $12$-gon. Here we consider a regular $12$-gon for simplicity.

The sides of the $12$-gon then correspond to the canonical closed loops of the closed surface which generate the fundamental group of $\Sigma_3$:
\begin{equation}\label{eq:g3_rels}
  \pi_1(\Sigma_3)=\langle \alpha_1,\beta_1,\cdots,\alpha_3,\beta_3\,|\,\prod_{i=1}^3\alpha_i \beta_i \alpha_i^{-1}\beta_i^{-1}=1 \rangle.
\end{equation}

Similar to the genus $2$ case, we can tile the hyperbolic plane by attaching copies of the standard $12$-gon
appropriately together, which then specifies a universal covering for $\Sigma_3$. We can choose one of the
standard $12$-gons as the fundamental domain of the covering. Such a tiling is plotted in Fig.\ref{fig:g3new}a.

The mapping class group of $\Sigma_3$ can be generated by the Dehn twists around the $\alpha$ and $\beta$ loops
plus the Dehn twists along $\gamma_1$ and $\gamma_2$ loops, where:
\begin{equation}
  \gamma_i=\alpha_{i+1}^{-1} \tilde \alpha_i,\qquad i=1,2
\end{equation}
and where $\tilde \alpha_i$ is the $\alpha_i$ but transported along $\beta_i$:
\begin{equation}
  \tilde\alpha_i=\beta_i \alpha_i \beta_i^{-1}
\end{equation}

To implement the Dehn twists, it suffices to find appropriate shearing maps to permute the qubits accordingly.
The subsequent step, which is the re-triangulation, has already been described in the general case in Section \ref{sec:reTriangulation}.
The representative maps for the Dehn twists along $\alpha$ and $\beta$ loops are essentially the same as
the ones described in Section \ref{sec:continuousMaps} for the $g = 2$ case.
As an example, Fig. \ref{fig:g3new}b shows how the fundamental domain is transformed by  $D^\ast_{\alpha_1}$ in this case.

The representative maps for Dehn twists along $\gamma_i$s are also straightforward generalizations of the map described in
Section \ref{sec:continuousMaps}; the trick is to choose a $12$-gon as the fundamental domain such that
it has the $\gamma_i$ loop as two of its sides and only the sides neighboring $\gamma_i$ transform non-trivially by the representative Dehn twist map.
Here we consider the Dehn twist along $\gamma_1$. The Dehn twist along $\gamma_2$ follows similarly.
The action of $D_{\gamma_1}$ on $\alpha_1$, $\alpha_2$, $\beta_1$ and $\beta_2$ follows from the expressions
in \eqref{eq:Dg} by replacing $\gamma$ with $\gamma_1$. Furthermore, it keeps $\alpha_3$ and $\beta_3$  invariant.
The loop $\delta_2=\beta_2 \beta_1$ also remains invariant under $D_{\gamma_1}$. Just
like \eqref{eq:modified_rel} for the $g=2$ case, we can rewrite the group relation in
\eqref{eq:g3_rels} as:
\begin{equation}
  \delta_2^{-1}  \tilde \alpha_2 \delta_2 \beta_1^{-1} \gamma_1 \beta_1 \gamma_1^{-1} \tilde \alpha_2^{-1}\alpha_3\beta_3\alpha_3^{-1}\beta_3=1.
\end{equation}
This form then suggests trading $\{\alpha_1,\beta_1\}$ with $\{\gamma_1,\beta_1\}$ and $\{\alpha_2,\beta_2\}$ with $\{\tilde \alpha_2, \delta_2\}$.
Fig. \ref{fig:g3new}c shows these new set of loops on the surface of $\Sigma_3$. If we cut $\Sigma_3$ along these loops,
we end up with the the shaded irregular $12$-gon shown in Fig. \ref{fig:g3new}d, which can be taken as the fundamental
domain of the covering map. Note that in this $12$-gon all sides remain invariant by $D_{\gamma_1}$ except the two sides
labeled by $\beta_1$. Moreover, the $\beta_1$ sides are neighboring the $\gamma_1$ sides as was the case in $g=2$.
Therefore a shearing map directly analogous to $D^\ast_\gamma$ in Section \ref{sec:continuousMaps} would work here as well.

\section{Logical gates on the hyperbolic $\mathbb{Z}_2$ surface code}\label{sec:surface}

\begin{figure*}[hbt]
\centering
 \includegraphics[width=2\columnwidth]{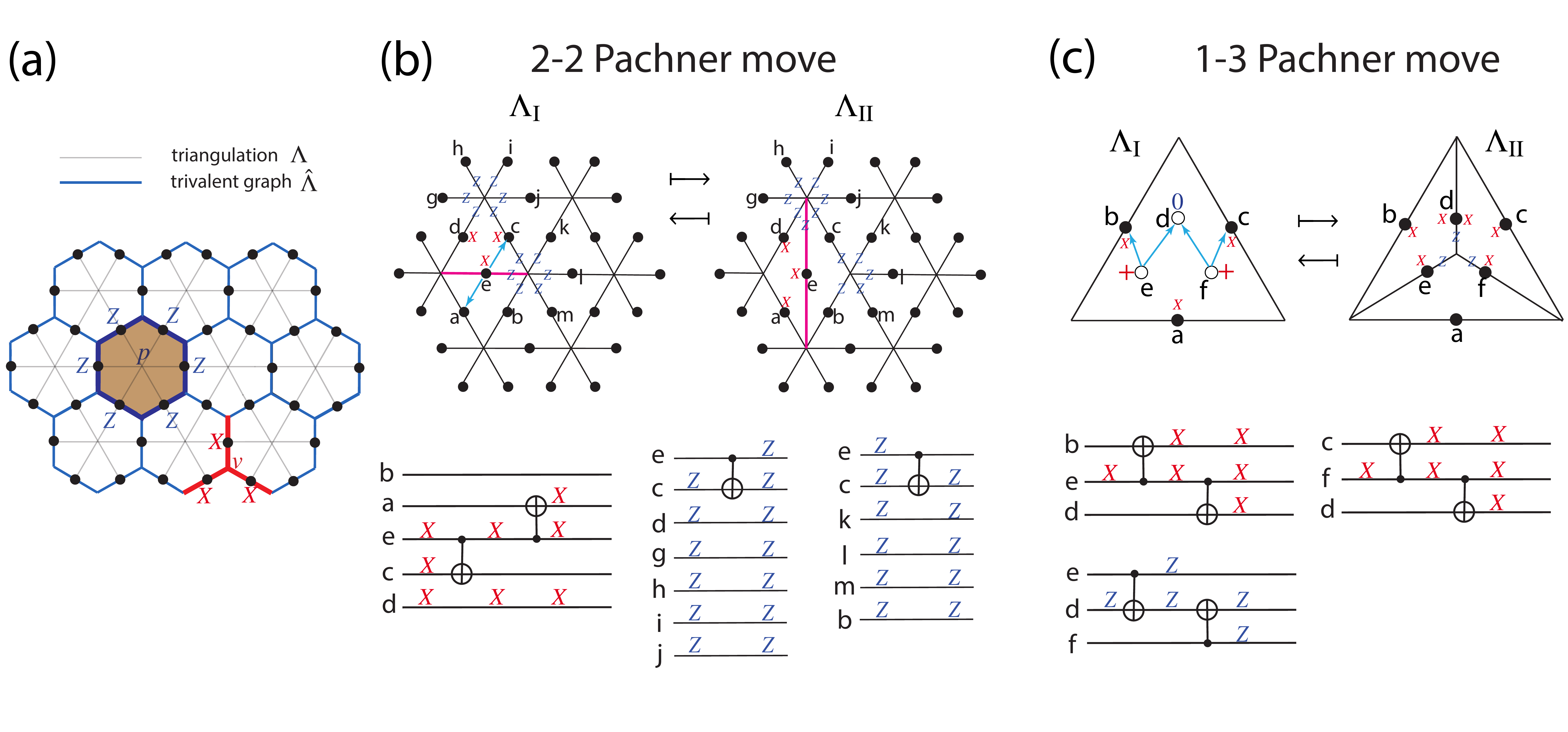}
  \caption{ (a) Definition of the surface code Hamiltonian on a trivalent graph $\hat{\Lambda}$. (b) Circuit implementation of the 2-2 Pachner move in the triangulation picture: $\Lambda_\text{I}\rightarrow \Lambda_\text{II}$. (c) Circuit implementation of the 1-3 Pachner move in the triangulation picture.}
  \label{fig:surface-code_retriangulation}
\end{figure*}

In this appendix, we discuss the specific case of applying our logical gate sets to the hyperbolic $\mathbb{Z}_2$ surface code, i.e. the code
obtained by placing the Kitaev $\mathbb{Z}_2$ toric code on a hyperbolic surface. This is the simplest abelian case of the
general Turaev-Viro code. In this case, the continuous shear map is exactly the same as what discussed for the general
Turaev-Viro code in Sec.~\ref{sec:continuousMaps} in the main text.  The retriangulation part also implements the same
sequence of 2-2 and 1-3 Pachner moves.  The implementation of the Pachner moves in terms unitary gates on the physical
qubits are even simpler than the Fibonacci code which we illustrated in the Fig.~\ref{fig:main_switching_algorithm} in the main text.
In the following, we show the concrete unitary circuits for the implementation of the 2-2 and 1-3 Pachner moves in the hyperbolic surface code.

We note that a number of previous discussions of the hyperbolic surface code considered the case of regular tilings
of the hyperbolic surface \cite{zemor2009,breuckmann2016,Breuckmann:2017hy}. Our protocols require us to consider
the code defined on more general triangulations.

As a specific situation of the general Levin-Wen Hamiltonian Eq.~\eqref{eq:levwen}, the $\mathbb{Z}_2$ surface code is described by the following
Hamiltonian defined on the trivalent graph $\hat{\Lambda}$ following the usual convention:
\begin{align}
H_{\mathbb{Z}_2}= - \sum_{\nu} Q_\nu - \sum_{p} B_p,
\end{align}
as illustrated by Fig.~\ref{fig:surface-code_retriangulation}(a). Here, the 3-leg vertex operator $Q_\nu$$=$$X_{\nu, 1} X_{\nu, 2} X_{\nu, 3}$
is a product of three Pauli-X operators due to the trivalent nature. The $j$-edge plaquette operator $B_p$$=$$Z_{\nu, 1} Z_{\nu, 2}\cdots Z_{\nu, j}$
is a product of $j$ Pauli-Z operators.  Note that the number of edges $j$ on a plaquette is in general arbitrary for an arbitrary trivalent
graph, but remains valid for a well-behaved graph to ensure the low-weight condition in the LDPC code family. For example, on a
honeycomb lattice,  one has $j=6$ as shown in Fig.~\ref{fig:surface-code_retriangulation}(a).   If we instead consider the dual
description of the hamiltonian on the original triangulation $\Lambda$, $Q_\nu$$=$$X_{\nu, 1} X_{\nu, 2} X_{\nu, 3}$ is located
on the triangle plaquettes on $\Lambda$, while  $B_p$$=$$Z_{\nu, 1} Z_{\nu, 2}\cdots Z_{\nu, j}$ is located on the $j$-leg vertices of $\Lambda$.

In the following, we explicitly demonstrate the circuit implementation of the 2-2 and 1-3 Pachner moves, focusing on the description
in the triangulation picture $\Lambda$ for clarity. For the convenience of our discussion, we will use the stabilizer formalism to
discuss the operator transformation in the Heisenberg picture.  In particular, we will use the following relations on the transformations
of the Pauli operators when conjugated by CNOTs:
\begin{align}\label{CNOT_relation}
\nonumber \text{CNOT} \ (X\otimes I) \ \text{CNOT} =& X\otimes X, \\
\nonumber \text{CNOT} \ (I\otimes X) \ \text{CNOT} =& I\otimes X, \\
\nonumber \text{CNOT} \ (Z\otimes I) \ \text{CNOT} =& Z\otimes I,\\
\text{CNOT} \ (I\otimes Z) \ \text{CNOT} =& Z\otimes Z,
\end{align}
where control is the first qubit and target the second one.
The inverse of the above transformation generated by conjugating with CNOTs again on both sides will also be used.

As shown in Fig.~\ref{fig:surface-code_retriangulation}(b), we want to implement a 2-2 Pachner move by flipping the edge
$e$ (pink) which re-divides the diamond $\Diamond_{abcd}$ into two triangles $\Delta_{ead}$ and $\Delta_{bce}$,  starting
from the original partition with triangles $\Delta_{ecd}$ and $\Delta_{abe}$. To achieve this, we apply two CNOTs indicated
by the blue arrow where the arrowhead points towards the target qubit and the tail points towards the control qubit.
In particular, the two CNOTs both start from qubit $e$ and point towards the qubit $c$ and $a$ respectively.
We start with a stabilizer $X_e X_c X_d$ in the original triangulation $\Lambda_\text{I}$ with eigenvalues $+1$.  As shown by the quantum circuit at the bottom of Fig.~\ref{fig:surface-code_retriangulation}(b),  the CNOT from $e$ to $c$ makes the following transformation of the stabilizer:
\begin{align}
\text{CNOT}_{ec}:X_e X_c X_d \longmapsto X_e X_d,
\end{align}
where the qubit $c$ is disentangled according to the inverse of the first relation in Eq.~\eqref{CNOT_relation}.  Next, the CNOT from $e$ and $a$ leads to
\begin{align}
\text{CNOT}_{ea}:X_e X_d \longmapsto X_e X_a X_d,
\end{align}
according to the first relation in Eq.~\eqref{CNOT_relation}.
Therefore, we get the new stabilizer $X_e X_a X_d$ with eigenvalue $+1$ on the new triangle $\Delta_{ead}$.  Note that the diamond stabilizer on $\Diamond_{abcd}$ is a product of the stabilizers on the two original triangles  $\Delta_{abe}$ and $\Delta_{ecd}$, i.e., $X_a X_b X_c X_d = (X_a X_b X_e)(X_e X_c X_d)$, having an eigenvalue $+1$ which is unchanged by the two CNOTs due to the second relations in Eq.~\eqref{CNOT_relation}.  At the same time, this diamond stabilizer can be also written as a product of the stabilizers on the two new triangles $\Delta_{bce}$ and $\Delta_{ead}$, i.e., $X_a X_b X_c X_d = (X_b X_c X_e)(X_e X_a X_d)$. Therefore, the other new triangle stabilizer $X_b X_c X_e$ should also have eigenvalue $+1$.

To check the consistency, one should also verify the corresponding transformation of the Z-stabilizers.  As one can see from the quantum circuit in Fig.~\ref{fig:surface-code_retriangulation}(b), the 6-body Z-stabilizer on top of the diamond is transformed by the CNOT from $e$ to $c$ according to the fourth relation in Eq.~\eqref{CNOT_relation} as:
 \begin{align}
\text{CNOT}_{ec}: Z_c Z_d Z_g Z_h Z_i Z_j \longmapsto  Z_e Z_c Z_d Z_g Z_h Z_i Z_j,
\end{align}
which entangles qubit $e$ and leads to a 7-body Z-stabilizer.   Similar transformation occurs for the Z-stabilizer at the bottom of the diamond.  Now the 6-body Z-stabilizer on the right of the diamond is transformed by the CNOT from $e$ to $c$ according to the inverse of the fourth relation in Eq.~\eqref{CNOT_relation} as:
\begin{align}
\text{CNOT}_{ec}: Z_e Z_c Z_k Z_l Z_m Z_b \longmapsto  Z_c Z_k Z_l Z_m Z_b ,
\end{align}
which disentangles the qubit $e$ and leads to a 5-body Z-stabilizer.  A similar transformation occurs for the Z-stabilizer on the left of the
diamond. Up to now, we have verified that these two CNOTs in Fig.~\ref{fig:surface-code_retriangulation}(b) accomplish the 2-2 Pachner move.

Now we consider the implementation of the 1-3 Pachner move by 4 CNOTs as shown in Fig.~\ref{fig:surface-code_retriangulation}(c).  We start with three ancilla qubits, with qubit $d$ initialized at $\ket{0}$, and qubit $e$ and $f$ initialized at $\ket{+}$. Note that $\ket{+}$ ($\ket{0}$) is an eigenstate of Pauli-X (Pauli-Z) with eigenvalue $+1$.
The two CNOTs starting from $e$ towards $b$ and $d$ create the triangular X-stabilizer on $\Delta_{bed}$ from a single Pauli-X on the ancilla $e$ with eigenvalue $+1$:
\begin{align}
\text{CNOT}_{eb}:X_e \longmapsto X_b X_e, \ \text{CNOT}_{ed}:X_b X_e \longmapsto X_b X_e X_d.
\end{align}
Similarly, the CNOTs acting from $f$ towards $c$ and $d$ create the triangular X-stabilizer on $\Delta_{bed}$ from a single Pauli-X on the ancilla $e$ with eigenvalue $+1$:
\begin{align}
\text{CNOT}_{fc}:X_f \longmapsto X_c X_f, \ \text{CNOT}_{fd}:X_c X_f \longmapsto X_c X_f X_d.
\end{align}

In addition, the two CNOTs from $e$ and $f$ towards $d$ makes sure the new Z-stabilizer is generated on the central vertex from the single Pauli-Z on qubit $d$ with eigenvalue $+1$:
\begin{align}
\text{CNOT}_{de}:Z_d \longmapsto Z_e Z_d, \ \text{CNOT}_{df}:Z_e Z_d \longmapsto Z_e Z_d Z_f.
\end{align}
Finally, one can verify the other Z-stabilizers are also transformed accordingly by the CNOTs.    Up to now, we have verified that the CNOTs in Fig.~\ref{fig:surface-code_retriangulation}(c) indeed implement the 1-3 Pachner move in the surface code.

Last but not least, we also note that the representation of the MCG on the hyperbolic surface code correspond to logical gates
in a subset of the Clifford group.  In the context of our scheme, this subset of logical gates can therefore be implemented
effectively instantaneously with long-range SWAP operations implementing the permutations and a constant depth local
quantum circuit implementing the subsequent retriangulation.

\bibliographystyle{plainnat}
\bibliography{mybib_merge.bib}

\begin{thebibliography}{58}
\providecommand{\natexlab}[1]{#1}
\providecommand{\url}[1]{\texttt{#1}}
\expandafter\ifx\csname urlstyle\endcsname\relax
  \providecommand{\doi}[1]{doi: #1}\else
  \providecommand{\doi}{doi: \begingroup \urlstyle{rm}\Url}\fi

\bibitem[Axline et~al.(2018)Axline, Burkhart, Pfaff, Zhang, Chou,
  Campagne-Ibarcq, Reinhold, Frunzio, Girvin, Jiang, Devoret, and
  Schoelkopf]{Axline:2017uq}
Christopher~J. Axline, Luke~D. Burkhart, Wolfgang Pfaff, Mengzhen Zhang, Kevin
  Chou, Philippe Campagne-Ibarcq, Philip Reinhold, Luigi Frunzio, S.~M. Girvin,
  Liang Jiang, M.~H. Devoret, and R.~J. Schoelkopf.
\newblock On-demand quantum state transfer and entanglement between remote
  microwave cavity memories.
\newblock \emph{Nature Physics}, 14\penalty0 (7):\penalty0 705--710, apr 2018.
\newblock \doi{10.1038/s41567-018-0115-y}.

\bibitem[Barkeshli and Freedman(2016)]{barkeshli2016mcg}
Maissam Barkeshli and Michael Freedman.
\newblock Modular transformations through sequences of topological charge
  projections.
\newblock \emph{Physical Review B}, 94\penalty0 (16), oct 2016.
\newblock \doi{10.1103/physrevb.94.165108}.

\bibitem[Barrett and Westbury(1996)]{barrett1996}
John~W. Barrett and Bruce~W. Westbury.
\newblock Invariants of piecewise-linear 3-manifolds.
\newblock \emph{Trans. Amer. Math. Soc.}, 348:\penalty0 3997--4022, 1996.
\newblock \doi{10.1090/S0002-9947-96-01660-1}.

\bibitem[Beverland et~al.(2016)Beverland, Buerschaper, Koenig, Pastawski,
  Preskill, and Sijher]{Beverland:2016bi}
Michael~E Beverland, Oliver Buerschaper, Robert Koenig, Fernando Pastawski,
  John Preskill, and Sumit Sijher.
\newblock {Protected gates for topological quantum field theories}.
\newblock \emph{Journal of Mathematical Physics}, 57\penalty0 (2):\penalty0
  022201--40, February 2016.
\newblock \doi{10.1063/1.4939783}.

\bibitem[Bogdanov and Teillaud(2013)]{Hyperbolic_Delanunay_closed_2013}
Mikhail Bogdanov and Monique Teillaud.
\newblock {Delaunay triangulations and cycles on closed hyperbolic surfaces}.
\newblock \emph{[Research Report] RR-8434, INRIA}, December 2013.

\bibitem[Bogdanov et~al.(2014)Bogdanov, Devillers, Teillaud, and
  Devillers]{Hyperbolic_Delanunay_2013}
Mikhail Bogdanov, Olivier Devillers, Monique Teillaud, and Olivier Devillers.
\newblock {Hyperbolic delaunay complexes and voronoi diagrams made practical}.
\newblock \emph{JoCG 5(1), 56--85}, December 2014.
\newblock \doi{10.1145/2462356.2462365}.

\bibitem[Bombin(2015)]{Bombin:2015hia}
Hector Bombin.
\newblock {Single-Shot Fault-Tolerant Quantum Error Correction}.
\newblock \emph{Physical Review X}, 5\penalty0 (3):\penalty0 181--26, September
  2015.
\newblock \doi{10.1103/PhysRevX.5.031043}.

\bibitem[Bonderson et~al.(2009)Bonderson, Freedman, and Nayak]{bonderson2009}
Parsa Bonderson, Michael Freedman, and Chetan Nayak.
\newblock Measurement-only topological quantum computation via anyonic
  interferometry.
\newblock \emph{Annals of Physics}, 324\penalty0 (4):\penalty0 787 -- 826,
  2009.
\newblock \doi{10.1016/j.aop.2008.09.009}.

\bibitem[Bonesteel and DiVincenzo(2012)]{Bonesteel:2012fl}
N~E Bonesteel and D~P DiVincenzo.
\newblock {Quantum circuits for measuring Levin-Wen operators}.
\newblock \emph{Physical Review B}, 86\penalty0 (16):\penalty0 165113, 2012.
\newblock \doi{10.1103/PhysRevB.86.165113}.

\bibitem[Bravyi and Hastings(2014)]{bravyi2014}
Sergey Bravyi and Matthew~B. Hastings.
\newblock Homological product codes.
\newblock \emph{Proc. of the 46th ACM Symposium on Theory of Computing (STOC
  2014)}, pages 273--282, 2014.
\newblock \doi{10.1145/2591796.2591870}.

\bibitem[Bravyi and Kitaev(2005)]{bravyi2005}
Sergey Bravyi and Alexei Kitaev.
\newblock Universal quantum computation with ideal clifford gates and noisy
  ancillas.
\newblock \emph{Phys. Rev. A}, 71:\penalty0 022316, Feb 2005.
\newblock \doi{10.1103/PhysRevA.71.022316}.

\bibitem[Bravyi and K{\"o}nig(2013)]{Bravyi:2013dx}
Sergey Bravyi and Robert K{\"o}nig.
\newblock {Classification of Topologically Protected Gates for Local Stabilizer
  Codes}.
\newblock \emph{Phys. Rev. Lett.}, 110\penalty0 (17):\penalty0 170503--5, April
  2013.
\newblock \doi{10.1103/PhysRevLett.110.170503}.

\bibitem[Bravyi et~al.(2010)Bravyi, Poulin, and Terhal]{bravyi2010tradeoffs}
Sergey Bravyi, David Poulin, and Barbara Terhal.
\newblock Tradeoffs for reliable quantum information storage in 2d systems.
\newblock \emph{Physical review letters}, 104\penalty0 (5):\penalty0 050503,
  2010.
\newblock \doi{10.1103/PhysRevLett.104.050503}.

\bibitem[Breuckmann and Terhal(2016)]{breuckmann2016}
Nikolas~P. Breuckmann and Barbara~M. Terhal.
\newblock Constructions and noise threshold of hyperbolic surface codes.
\newblock \emph{IEEE Transactions on Information Theory}, 62, 2016.
\newblock \doi{10.1109/TIT.2016.2555700}.

\bibitem[Breuckmann et~al.(2017)Breuckmann, Vuillot, Campbell, Krishna, and
  Terhal]{Breuckmann:2017hy}
Nikolas~P Breuckmann, Christophe Vuillot, Earl Campbell, Anirudh Krishna, and
  Barbara~M Terhal.
\newblock {Hyperbolic and semi-hyperbolic surface codes for quantum storage}.
\newblock \emph{Quantum Science and Technology}, 2\penalty0 (3):\penalty0
  035007--21, August 2017.
\newblock \doi{10.1088/2058-9565/aa7d3b}.

\bibitem[Burton et~al.(2017)Burton, Brell, and Flammia]{Burton:2017gr}
Simon Burton, Courtney~G Brell, and Steven~T Flammia.
\newblock {Classical simulation of quantum error correction in a Fibonacci
  anyon code}.
\newblock \emph{Physical Review A}, 95\penalty0 (2):\penalty0 580--10, February
  2017.
\newblock \doi{10.1103/PhysRevA.95.022309}.

\bibitem[Calderbank and Shor(1996)]{calderbank1996}
A.~R. Calderbank and Peter~W. Shor.
\newblock Good quantum error-correcting codes exist.
\newblock \emph{Phys. Rev. A}, 54:\penalty0 1098--1105, Aug 1996.
\newblock \doi{10.1103/PhysRevA.54.1098}.

\bibitem[Campagne-Ibarcq et~al.(2018)Campagne-Ibarcq, Zalys-Geller, Narla,
  Shankar, Reinhold, Burkhart, Axline, Pfaff, Frunzio, Schoelkopf, and
  Devoret]{CampagneIbarcq:2017wq}
P.~Campagne-Ibarcq, E.~Zalys-Geller, A.~Narla, S.~Shankar, P.~Reinhold,
  L.~Burkhart, C.~Axline, W.~Pfaff, L.~Frunzio, R.~J. Schoelkopf, and M.~H.
  Devoret.
\newblock Deterministic remote entanglement of superconducting circuits through
  microwave two-photon transitions.
\newblock \emph{Phys. Rev. Lett.}, 120:\penalty0 200501, May 2018.
\newblock \doi{10.1103/PhysRevLett.120.200501}.

\bibitem[Chou et~al.(2018)Chou, Blumoff, Wang, {Reinhold, P. C.}, Axline, Gao,
  Frunzio, Devoret, Jiang, and Schoelkopf]{Chou:2018vz}
K~S Chou, J~Z Blumoff, C~S Wang, {Reinhold, P. C.}, C~J Axline, Y~Y Gao,
  L~Frunzio, M~H Devoret, Liang Jiang, and R~J Schoelkopf.
\newblock {Deterministic teleportation of a quantum gate between two logical
  qubits}.
\newblock January 2018.
\newblock \doi{10.1038/s41586-018-0470-y}.

\bibitem[Comparat and Pillet(2010)]{Comparat:2010cb}
Daniel Comparat and Pierre Pillet.
\newblock {Dipole blockade in a cold Rydberg atomic sample [Invited]}.
\newblock \emph{Journal of the Optical Society of America B}, 27\penalty0
  (6):\penalty0 A208--A232, June 2010.
\newblock \doi{10.1364/JOSAB.27.00A208}.

\bibitem[Dauphinais and Poulin(2017)]{Dauphinais:2017bz}
Guillaume Dauphinais and David Poulin.
\newblock {Fault-Tolerant Quantum Error Correction for non-Abelian Anyons}.
\newblock \emph{Communications in Mathematical Physics}, 355\penalty0
  (2):\penalty0 519--560, July 2017.
\newblock \doi{10.1007/s00220-017-2923-9}.

\bibitem[Dehn(2012)]{dehn2012papers}
Max Dehn.
\newblock \emph{Papers on group theory and topology}.
\newblock Springer Science \& Business Media, 2012.
\newblock \doi{10.1007/978-1-4612-4668-8}.

\bibitem[Farb and Margalit(2011)]{farb2011primer}
Benson Farb and Dan Margalit.
\newblock \emph{A primer on mapping class groups (pms-49)}.
\newblock Princeton University Press, 2011.
\newblock \doi{10.23943/princeton/9780691147949.001.0001}.

\bibitem[Feng(2015)]{feng2015non}
Weibo Feng.
\newblock Non-abelian quantum error correction.
\newblock \emph{Ph.D. Thesis, The Florida State University}, 2015.

\bibitem[Freedman et~al.(2002{\natexlab{a}})Freedman, Larsen, and
  Wang]{Freedman_Larsen_wang_2002}
M~H Freedman, M~Larsen, and Z~H Wang.
\newblock {A modular functor which is universal for quantum computation}.
\newblock \emph{Communications in Mathematical Physics}, 227\penalty0
  (3):\penalty0 605--622, June 2002{\natexlab{a}}.
\newblock \doi{10.1007/s002200200645}.

\bibitem[Freedman et~al.(2002{\natexlab{b}})Freedman, Meyer, and
  Luo]{freedman2002z2}
Michael~H Freedman, David~A Meyer, and Feng Luo.
\newblock Z2-systolic freedom and quantum codes.
\newblock \emph{Mathematics of quantum computation, Chapman \& Hall/CRC}, pages
  287--320, 2002{\natexlab{b}}.

\bibitem[Fricke and Klein(1897)]{fricke1897vorlesungen}
Robert Fricke and Felix Klein.
\newblock \emph{Vorlesungen uber die Theorie der automorphen Funktionen}.
\newblock Johnson Reprint, 1897.
\newblock \doi{10.1007/BF01699777}.

\bibitem[Gottesman(2014)]{Gottesman:2014ug}
Daniel Gottesman.
\newblock {Fault-Tolerant Quantum Computation with Constant Overhead}.
\newblock \emph{Quantum Information {\&} Computation}, 14\penalty0
  (15-16):\penalty0 1338--1371, November 2014.

\bibitem[Guth and Lubotzky(2014)]{Guth:2014cj}
Larry Guth and Alexander Lubotzky.
\newblock {Quantum error correcting codes and 4-dimensional arithmetic
  hyperbolic manifolds}.
\newblock \emph{Journal of Mathematical Physics}, 55\penalty0 (8):\penalty0
  082202, August 2014.
\newblock \doi{10.1063/1.4891487}.

\bibitem[Herrera-Mart{\'\i} et~al.(2010)Herrera-Mart{\'\i}, Fowler, Jennings,
  and Rudolph]{HerreraMarti:2010cu}
David~A Herrera-Mart{\'\i}, Austin~G Fowler, David Jennings, and Terry Rudolph.
\newblock {Photonic implementation for the topological cluster-state quantum
  computer}.
\newblock \emph{Physical Review A}, 82\penalty0 (3):\penalty0 032332--6,
  September 2010.
\newblock \doi{10.1103/PhysRevA.82.032332}.

\bibitem[Hjelle and D{\ae}hlen()]{Hjelle:2006wr}
{\O}~Hjelle and M~D{\ae}hlen.
\newblock \emph{{Triangulations and applications}}.
\newblock Springer; 2006 edition (September 19, 2006).
\newblock \doi{10.1007/3-540-33261-8}.

\bibitem[Keen(1966)]{keen1966canonical}
Linda Keen.
\newblock Canonical polygons for finitely generated fuchsian groups.
\newblock \emph{Acta Mathematica}, 115\penalty0 (1):\penalty0 1--16, 1966.
\newblock \doi{10.1007/BF02392200}.

\bibitem[Kitaev(2003)]{kitaev2003}
A.Yu. Kitaev.
\newblock Fault-tolerant quantum computation by anyons.
\newblock \emph{Annals Phys.}, 303:\penalty0 2--30, 2003.
\newblock \doi{10.1016/S0003-4916(02)00018-0}.

\bibitem[Koenig et~al.(2010)Koenig, Kuperberg, and Reichardt]{Koenig:2010do}
Robert Koenig, Greg Kuperberg, and Ben~W Reichardt.
\newblock {Quantum computation with Turaev-Viro codes}.
\newblock \emph{Annals of Physics}, 325\penalty0 (12):\penalty0 2707--2749,
  December 2010.
\newblock \doi{10.1016/j.aop.2010.08.001}.

\bibitem[Koll{\'a}r et~al.(2019)Koll{\'a}r, Fitzpatrick, and
  Houck]{cKollar:2018vc}
Alicia~J. Koll{\'a}r, Mattias Fitzpatrick, and Andrew~A. Houck.
\newblock Hyperbolic lattices in circuit quantum electrodynamics.
\newblock \emph{Nature}, 571\penalty0 (7763):\penalty0 45--50, 2019.
\newblock \doi{10.1038/s41586-019-1348-3}.

\bibitem[Kovalev and Pryadko(2013)]{kovalev2013}
Alexey~A. Kovalev and Leonid~P. Pryadko.
\newblock Fault tolerance of quantum low-density parity check codes with
  sublinear distance scaling.
\newblock \emph{Phys. Rev. A}, 87:\penalty0 020304, Feb 2013.
\newblock \doi{10.1103/PhysRevA.87.020304}.

\bibitem[Kurpiers et~al.(2018)Kurpiers, Magnard, Walter, Royer, Pechal,
  Heinsoo, Salath{\'e}, Akin, Storz, Besse, et~al.]{Kurpiers:2017ub}
Philipp Kurpiers, Paul Magnard, Theo Walter, Baptiste Royer, Marek Pechal,
  Johannes Heinsoo, Yves Salath{\'e}, Abdulkadir Akin, Simon Storz, J-C Besse,
  et~al.
\newblock Deterministic quantum state transfer and remote entanglement using
  microwave photons.
\newblock \emph{Nature}, 558\penalty0 (7709):\penalty0 264, 2018.
\newblock \doi{10.1038/s41586-018-0195-y}.

\bibitem[Lavasani and Barkeshli(2018)]{lavasani2018}
Ali Lavasani and Maissam Barkeshli.
\newblock Low overhead clifford gates from joint measurements in surface,
  color, and hyperbolic codes.
\newblock 2018.
\newblock \doi{10.1103/PhysRevA.98.052319}.

\bibitem[Lekitsch et~al.(2017)Lekitsch, Weidt, Fowler, M{\o}lmer, Devitt,
  Wunderlich, and Hensinger]{Lekitsch:2015ua}
Bjoern Lekitsch, Sebastian Weidt, Austin~G. Fowler, Klaus M{\o}lmer, Simon~J.
  Devitt, Christof Wunderlich, and Winfried~K. Hensinger.
\newblock Blueprint for a microwave trapped ion quantum computer.
\newblock \emph{Science Advances}, 3\penalty0 (2), 2017.
\newblock \doi{10.1126/sciadv.1601540}.

\bibitem[Leverrier et~al.(2015)Leverrier, Tillich, and
  Z{\'e}mor]{Leverrier:2015ju}
Anthony Leverrier, Jean-Pierre Tillich, and Gilles Z{\'e}mor.
\newblock Quantum expander codes.
\newblock pages 810--824, 2015.
\newblock \doi{10.1109/FOCS.2015.55}.

\bibitem[Levin and Wen(2005)]{levin2005}
Michael~A. Levin and Xiao-Gang Wen.
\newblock String-net condensation: A physical mechanism for topological phases.
\newblock \emph{Phys. Rev. B}, 71:\penalty0 045110, Jan 2005.
\newblock \doi{10.1103/PhysRevB.71.045110}.

\bibitem[Linke et~al.(2017)Linke, Maslov, Roetteler, Debnath, Figgatt,
  Landsman, Wright, and Monroe]{Linke:2017bz}
N~M Linke, D~Maslov, M~Roetteler, S~Debnath, C~Figgatt, K~A Landsman, K~Wright,
  and C~Monroe.
\newblock {Experimental Comparison of Two Quantum Computing Architectures}.
\newblock \emph{PNAS 13, 3305--3310}, February 2017.
\newblock \doi{10.1073/pnas.1618020114}.

\bibitem[Maller et~al.(2015)Maller, Lichtman, Xia, Sun, Piotrowicz, Carr,
  Isenhower, and Saffman]{Maller:2015is}
K~M Maller, M~T Lichtman, T~Xia, Y~Sun, M~J Piotrowicz, A~W Carr, L~Isenhower,
  and M~Saffman.
\newblock {Rydberg-blockade controlled-not gate and entanglement in a
  two-dimensional array of neutral-atom qubits}.
\newblock \emph{Physical Review A}, 92\penalty0 (2):\penalty0 022336--12,
  August 2015.
\newblock \doi{10.1103/PhysRevA.92.022336}.

\bibitem[Munkres(1960)]{munkres1960obstructions}
James Munkres.
\newblock Obstructions to the smoothing of piecewise-differentiable
  homeomorphisms.
\newblock \emph{Annals of Mathematics}, pages 521--554, 1960.
\newblock \doi{10.2307/1970228}.

\bibitem[Nayak et~al.(2008)Nayak, Simon, Stern, Freedman, and Sarma]{nayak2008}
Chetan Nayak, Steven~H. Simon, Ady Stern, Michael Freedman, and Sankar~Das
  Sarma.
\newblock Non-abelian anyons and topological quantum computation.
\newblock 80:\penalty0 1083, 2008.
\newblock \doi{10.1103/RevModPhys.80.1083}.

\bibitem[Nielsen(1927)]{nielsen1927untersuchungen}
Jakob Nielsen.
\newblock Untersuchungen zur topologie der geschlossenen zweiseitigen
  fl{\"a}chen.
\newblock \emph{Acta Mathematica}, 50\penalty0 (1):\penalty0 189--358, 1927.
\newblock \doi{10.1007/BF02421324}.

\bibitem[Nielsen and Chuang(2010)]{nielsen_chuang_2010}
Michael~A. Nielsen and Isaac~L. Chuang.
\newblock \emph{Quantum Computation and Quantum Information: 10th Anniversary
  Edition}.
\newblock Cambridge University Press, 2010.
\newblock \doi{10.1017/CBO9780511976667}.

\bibitem[Paetznick and Reichardt(2013)]{Paetznick:2013fu}
Adam Paetznick and Ben~W Reichardt.
\newblock {Universal Fault-Tolerant Quantum Computation with Only Transversal
  Gates and Error Correction}.
\newblock \emph{Phys. Rev. Lett.}, 111\penalty0 (9):\penalty0 090505--5, August
  2013.
\newblock \doi{10.1103/PhysRevLett.111.090505}.

\bibitem[Pichler et~al.(2016)Pichler, Zhu, Seif, Zoller, and
  Hafezi]{Pichler:2016ec}
Hannes Pichler, Guanyu Zhu, Alireza Seif, Peter Zoller, and Mohammad Hafezi.
\newblock {Measurement Protocol for the Entanglement Spectrum of Cold Atoms}.
\newblock \emph{Physical Review X}, 6\penalty0 (4):\penalty0 041033--12,
  November 2016.
\newblock \doi{10.1103/PhysRevX.6.041033}.

\bibitem[Saffman et~al.(2010)Saffman, Walker, and M{\o}lmer]{Saffman:2010ky}
M~Saffman, T~G Walker, and K~M{\o}lmer.
\newblock {Quantum information with Rydberg atoms}.
\newblock \emph{Rev. Mod. Phys.}, 82\penalty0 (3):\penalty0 2313--2363, August
  2010.
\newblock \doi{10.1103/RevModPhys.82.2313}.

\bibitem[Terhal(2015)]{Terhal:2015ks}
Barbara~M Terhal.
\newblock {Quantum error correction for quantum memories}.
\newblock \emph{Rev. Mod. Phys.}, 87\penalty0 (2):\penalty0 307--346, April
  2015.
\newblock \doi{10.1103/RevModPhys.87.307}.

\bibitem[Tillich and Z{\'e}mor(2014)]{Tilich2014}
J.-P Tillich and Gilles Z{\'e}mor.
\newblock Quantum ldpc codes with positive rate and minimum distance
  proportional to the square root of the blocklength.
\newblock \emph{Information Theory, IEEE Transactions on}, 60:\penalty0
  1193--1202, 02 2014.
\newblock \doi{10.1109/TIT.2013.2292061}.

\bibitem[Turaev and Viro(1992)]{turaev1992}
V.G. Turaev and O.Y. Viro.
\newblock State sum invariants of 3-manifolds and quantum 6j-symbols.
\newblock \emph{Topology}, 31:\penalty0 865--902, 1992.
\newblock \doi{10.1016/0040-9383(92)90015-A}.

\bibitem[Wang(2010)]{zhwang2010}
Zhenghan Wang.
\newblock \emph{Topological Quantum Computation}.
\newblock American Mathematics Society, 2010.

\bibitem[Z{\'e}mor(2009)]{zemor2009}
Gilles Z{\'e}mor.
\newblock On cayley graphs, surface codes, and the limits of homological coding
  for quantum error correction.
\newblock In Yeow~Meng Chee, Chao Li, San Ling, Huaxiong Wang, and Chaoping
  Xing, editors, \emph{Coding and Cryptology}, pages 259--273, Berlin,
  Heidelberg, 2009. Springer Berlin Heidelberg.
\newblock ISBN 978-3-642-01877-0.
\newblock \doi{10.1007/978-3-642-01877-0_21}.

\bibitem[Zhu et~al.(2017)Zhu, Hafezi, and Barkeshli]{Zhu:2017tr}
Guanyu Zhu, Mohammad Hafezi, and Maissam Barkeshli.
\newblock {Quantum Origami: Transversal Gates for Quantum Computation and
  Measurement of Topological Order}.
\newblock \emph{arXiv:1711.05752}, November 2017.

\bibitem[Zhu et~al.(2018{\natexlab{a}})Zhu, Lavasani, and
  Barkeshli]{Zhu:2018CodeLong}
Guanyu Zhu, Ali Lavasani, and Maissam Barkeshli.
\newblock Instantaneous braids and dehn twists in topologically ordered states.
\newblock \emph{arXiv preprint arXiv:1806.06078}, 2018{\natexlab{a}}.

\bibitem[Zhu et~al.(2018{\natexlab{b}})Zhu, Lavasani, and Barkeshli]{zhu2018}
Guanyu Zhu, Ali Lavasani, and Maissam Barkeshli.
\newblock Fast universal logical gates on topologically encoded qubits at
  arbitrarily large code distances.
\newblock \emph{arXiv preprint arXiv:1806.02358}, 2018{\natexlab{b}}.

\end{thebibliography}
\end{document}